\documentclass[12pt]{article}
\usepackage{graphicx}
\usepackage{epsfig}
\usepackage{epstopdf}
\usepackage{amsmath}
\usepackage{amsfonts}
\usepackage{amssymb}
\usepackage{ccaption}
\usepackage{setspace}
\usepackage[usenames]{color}

\usepackage[
      colorlinks=true,
      linkcolor=black,  
      citecolor=black,
      urlcolor=black,    
      filecolor=blue,     
      pdfstartview=FitV,
      pdftitle={On Universality of Ergoregion Mergers},
        pdfauthor={Elvang, Figueras, Horowitz, Hubeny, Rangamani},
        pdfsubject={ergoregions},
        pdfkeywords={ergoregions, GR, black holes, saturn},
        pdfpagemode=None,
        bookmarksopen=true    
      ]{hyperref}

  \captionnamefont{\bfseries}
  \captiontitlefont{\small\sffamily}
  \captiondelim{: }
  \hangcaption

\newcommand{\rom}[1]{\mathrm{#1}}

\newcommand{\beq}{\begin{equation}}
\newcommand{\eeq}{\end{equation}}
\newcommand{\be}{\begin{equation}}
\newcommand{\ee}{\end{equation}}
\newcommand{\beqa}{\begin{eqnarray}}
\newcommand{\eeqa}{\end{eqnarray}}
\newcommand{\beqar}{\begin{eqnarray*}}
\newcommand{\eeqar}{\end{eqnarray*}}
\newcommand{\bea}{\begin{eqnarray}}
\newcommand{\eea}{\end{eqnarray}}

\newcommand{\p}{\partial}

\newcommand{\bc}{\bar{c}}

\newcommand{\ka}{\kappa}

\def\sec#1{Section \ref{#1}}
\def\fig#1{Fig.\,\ref{#1}}
\def\req#1{(\ref{#1})}
\def\App#1{Appendix \ref{#1}}

\def\a{\alpha}

\def\p{\partial}

\def\Om{\Omega}

\def\({\left(}
\def\){\right)}

\def\CB{{\cal B}}

\def\CO{{\cal O}}

\def\CR{{\cal R}}

\def\Sp{{\bf S}}

\def\A5S5{{\rm AdS}_5 \times \S^5}

\def\half{{\frac{1}{2}}}

\def\p{\partial}

\def\half{{\frac{1}{2}}}

\def\p{\partial}

\def\slope{\sigma}

\def\G{\gamma}

\newcommand{\labell}[1]{\label{#1}} %\qquad_{#1}} %{\label{#1}}
\newcommand{\reef}[1]{(\ref{#1})}

\newcommand{\Tr}{\textrm{Tr}}
\newcommand{\diag}{\textrm{diag}}
\numberwithin{equation}{section}
\begin{document}

\setlength{\unitlength}{1mm}

\begin{titlepage}

\begin{flushright}
DCPT-08/55\\
MIT-CTP-3986\\
UUITP-23/08
\end{flushright}
\vspace{1cm}

\vspace{-8mm}

\begin{center}
{\bf \Large On Universality in Ergoregion Mergers}
\end{center}

\begin{center}
Henriette Elvang$^{a}$\footnote{On leave of absence from Uppsala University.},
Pau Figueras$^{b}$,
Gary T.~Horowitz$^{c}$,\\[2mm]
Veronika E.~Hubeny$^{b}$ and
Mukund Rangamani$^{b}$

\vspace{.4cm}
{\small {\textit{$^{a}$Institute for Advanced Study,}}\\
{\small \textit{Einstein Drive, Princeton, NJ 08540, USA}}} \\
\vspace{2mm}
{\small \textit{$^{b}$Centre for Particle Theory \& Department of Mathematical Sciences,}}\\
{\small \textit{Science Laboratories, South Road, Durham DH1 3LE, United Kingdom}} \\
\vspace{2mm}
 {\small \textit{$^{c}$Department of Physics, UCSB,}}\\
{\small \textit{Santa Barbara, CA 93106, USA}}

\vspace*{0.3cm}
{\small {\tt elvang@ias.edu, pau.figueras@durham.ac.uk, gary@physics.ucsb.edu, veronika.hubeny@durham.ac.uk, mukund.rangamani@durham.ac.uk}}
\end{center}

%\vspace{4mm}

\begin{abstract}

We study mergers of ergoregions in $d+1$-dimensional vacuum gravity. At the merger point, where the ergosurfaces bounding each ergoregion just touch, solutions exhibit universal behavior when there is rotation only in one plane: the angle between the merging ergosurfaces depends only on the symmetries of the solution, not on any other details of the configuration. 
We show that universality follows from the fact that the relevant component of Einstein's equation reduces to Laplace equation at the point of merger. Thus ergoregion mergers mimic mergers of Newtonian equipotentials and have similar universal behavior. For solutions with rotation in more than one plane, universality is lost. We demonstrate universality and non-universality in several explicit examples.

\end{abstract}

\end{titlepage}

\setstretch{0.5}
\tableofcontents
\setstretch{1.1}

\newpage
%____________________________________________
\section{Introduction}
\label{intro}
%____________________________________________

In general relativity, rotating objects influence their surroundings by rotational dragging. The dragging effect can be so strong that everything gets `swept along' by the spacetime: in stationary spacetimes, an ergoregion is defined as a region of spacetime where it is impossible for any observer to remain at rest with respect to a static asymptotic observer. The boundary of the ergoregion, the ergosurface, can simply be defined as the locus of spacetime points where the asymptotic timelike Killing field becomes null, i.e.~where $G_{tt}=0$. The most familiar example is the Kerr black hole, whose ergoregion is bounded by a 2-sphere ergosurface surrounding the event horizon.
  
Physics of ergoregions has been well-studied in four dimensions, in particular in the context of the Penrose process \cite{Penrose:1969pc} for energy extraction from a rotating black hole. Superradiant scattering \cite{Misner,Zeldovich,Starobinskii} also relies on the existence of an ergoregion, and studies thereof even include stringy models for the microscopic process --- recent progress was reported in \cite{Dias:2007nj} (see also references therein and \cite{Chowdhury:2008bd}). However, there are only a few formal results concerning the properties of the ergosurface itself. One such result is Hajicek's theorem \cite{Hajicek:1973fk}: in four dimensions ergosurfaces must either touch the horizon or hit the black hole singularity at the fixed points of the rotational isometry. For instance, for the Kerr black hole, the ergosurface touches the $\mathbf{S}^2$ horizon at the poles.
However, this result does not extend beyond four dimensions because the horizons of  higher-dimensional black holes need not contain the fixed point of the rotational symmetry. An example is the black ring  \cite{Emparan:2001wn} for which the $\mathbf{S}^1$ of the ring is finite size everywhere on the horizon. Its ergosurface has topology $\mathbf{S}^2 \times \mathbf{S}^1$ and touches the horizon nowhere.

In this paper we study ergosurfaces in stationary vacuum spacetimes in general relativity, with focus on ergosurface topology change. This occurs when two disjoint ergoregions merge to form one large ergoregion, or when a single ergoregion deforms to change the topology of its boundary.  Rather than discussing dynamic mergers, we restrict our study to mergers within families of stationary solutions.
We derive general results revealing universal properties associated with ergoregion mergers and we illustrate these properties in examples of exact solutions for 5-dimensional vacuum black hole spacetimes.

Over the last few years, vacuum general relativity in higher dimensions has revealed intriguing new surprises, challenging the conventional paradigms built on four dimensional gravity. The discovery of the vacuum black ring solution in five dimensions in \cite{Emparan:2001wn} has spurred the investigation of higher dimensional solutions to Einstein's equation. In particular, building upon the class of Weyl solutions \cite{Emparan:2001wk,Harmark:2004rm} and exploiting the integrable nature of Einstein's equation with adequate symmetries \cite{Belinsky:1971nt,Belinsky:1979mh}, many novel vacuum black hole solutions have been constructed. For example, we now have a vacuum black ring in five dimensions that spins in both independent 2-planes \cite{Pomeransky:2006bd,Mishima:2005id,Figueras:2005zp,Tomizawa:2005wv}. Also interesting are the five-dimensional multi-black hole solutions, such as Black Saturn \cite{Elvang:2007rd} (a rotating black ring with a spinning black hole at its
  centre), the di-rings \cite{Iguchi:2007is,Evslin:2007fv} (two black rings in the same plane), and the bi-rings \cite{Izumi:2007qx, Elvang:2007hs} (two black rings in orthogonal planes). In these solutions, the angular momentum of the black ring(s) keeps the configuration in a balanced equilibrium. Thus the solutions are free of any naked singularities, they are regular on and outside the horizons.
An overview of higher dimensional black holes and their properties can be found in the recent reviews \cite{Emparan:2006mm,Emparan:2008eg, Obers:2008pj}.

\begin{figure}[t!]
\centering{\includegraphics[width=4in]{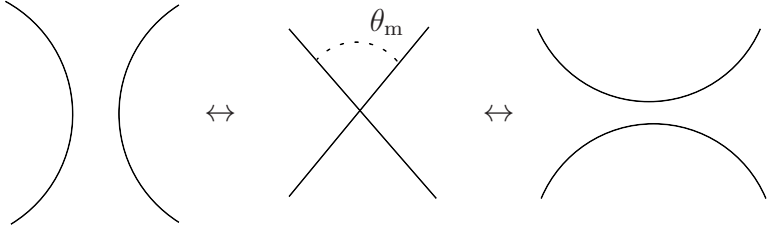}}
\begin{picture}(0,0)(0,0)
\put(-39,14){$\leftrightarrow$}
\put(-76,14){$\leftrightarrow$}
\put(-54,26){$\theta_\rom{m}$}
\end{picture}
\caption{Sketch of topology change in merger of ergosurface locii. The merger angle $\theta_\rom{m}$ is the angle between the tangents of the surfaces at the merger point.}
\label{fig:topch}
\end{figure}

Multi-black hole spacetimes, such as Black Saturn, provide excellent laboratories for studying ergoregion mergers. To exemplify the novel feature afforded by multiple ergoregions, let us consider a configuration with two rotating black holes, both with non-vanishing angular velocity. Each black hole has an ergoregion surrounding it. 
If the black holes are far apart, the gravitational interaction between the two constituents is weak, and we expect the respective ergoregions to be confined to the vicinity of the individual black hole.  However, as we dial the parameters of the solution to move the black holes closer together, the increase in the gravitational interaction will cause the ergoregions to distort. When the black holes get close enough, the distortion could be so large that the two ergoregions join into a single connected ergoregion, in spite of the horizons remaining disjoint. 
The transition\footnote{The phenomenon of merging ergosurfaces has been noticed before in the context of accretion of matter by black holes \cite{Ansorg:2000rn,Ansorg:2005bq}, four dimensional Ernst solutions \cite{Chrusciel:2006se}, and Kaluza-Klein black holes \cite{Matsuno:2008fn}.} from disjoint ergoregions to a single connected ergoregion passes through a special configuration where the two ergoregions just touch each other, as sketched in \fig{fig:topch}. This point is guaranteed to exist by continuity.  It is in fact easy to show that the merger point corresponds to  a static null geodesic in the spacetime, i.e.~a null geodesic whose tangent is just along the asymptotic time translational Killing field; generically, stationary spacetimes do not admit such static null geodesics.

%figure

An intriguing feature of the merger configurations is that they provide interesting examples of geometric surfaces undergoing  topology change in vacuum general relativity. For example, in the prototypical case of the Black Saturn spacetime, the black hole at the center has an ergosurface with topology $\Sp^3$, while the black ring has an ergosurface of topology $\Sp^2 \times \Sp^1$. After the merger, the ergosurface topology of the system is a single $\Sp^3$ surrounding both the hole and the ring. Another example is the doubly spinning black ring whose ergosurface can change topology from $\Sp^2 \times \Sp^1$ to $\Sp^3$ as the ring radius shrinks and the $\Sp^2$ angular momentum grows. A new feature is the appearance of an inner ergosurface of $\Sp^3$ topology which is inside the outer $\Sp^3$ (but outside the event horizon), and shields the symmetry point at the center of the ring from being part of the ergoregion.

{For} comparison, consider the four dimensional Kerr solution.\footnote{For a review of ergoregions of the Kerr black hole, see \cite{Visser:2007fj}, and for studies of the geometry \cite{Pelavas:2000za,Jacobson:2008nx}.} The usual ergosurface lies outside the event horizon, but the equation $G_{tt}=0$, which characterizes the ergosurface, also has a second solution which gives an ``inner ergosurface". It lies inside the inner horizon, touches it at the poles and touches the singularity at the equator. In the extremal limit, the inner and outer horizon coincide, and the inner and outer ergoregions touch at the poles of horizon. The nakedly singular over-spinning Kerr solution has a toroidal ergoregion surrounding the singularity. Thus the Kerr solution also provides an example of ergosurface topology change, but it involves ergosurfaces separated by a horizon and requires going to a nakedly singular limit. Both inconveniences can be avoided for ergoregion mergers in higher dimensional black hole spacetimes.
  
Given that ergosurfaces can merge, an obvious question concerns the behaviour of the local 
geometry near the merger point which exists in the subfamily of solutions tuned to the point where the ergosurfaces just touch. Specifically, we are interested in the angle $\theta_\rom{m}$ between the merging surfaces at the merger point, as indicated in \fig{fig:topch}. 
We can define this angle $\theta_\rom{m}$ in a coordinate-independent way using the scalar product between the unit normals to the ergosurfaces at the merger point.\footnote{This is a well-defined procedure since the ergosurfaces are generically (and in particular in a neighbourhood of the merger point) timelike, and therefore unique spacelike unit normals exist.  Operationally, we can perform the calculation conveniently in  Riemann normal coordinates at the merger point.}
In our work we impose symmetries, so that the metric only depends on two coordinates, say $(z,\rho)$, at the merger point. We can choose $(z,\rho)$ such that $G_{z\rho}=0$. This allow us to compute $\theta_\rom{m}$ directly as the angle between the tangents to the ergosurface locii in the $(z,\rho)$-plane.

Consider first $d+1$-dimensional vacuum solutions of the Weyl class. There are $d-1$ commuting Killing vectors, one of which will be the stationary time. The solution depends  on two coordinates $(z,\rho)$ with $\rho \ge 0$ and $-\infty < z < \infty$. Let us restrict ourselves to solutions with rotation only in a single plane (``singly spinning''). It then turns out that there are only two possibilities for the merger angle: either
\bea
  \label{Weylcase}
  \theta_\rom{m} = 2 \, \text{arccot}\sqrt{2}
   ~~~~~\text{or}~~~~~~
    \theta_\rom{m} = \frac{\pi}{2} \, .
\eea 
The two cases are distinguished by whether the merger point lies at $\rho = 0$ or $\rho > 0$. This will be described in detail in \sec{weylergo} and illustrated further in the examples in \sec{univex}.
In the first case of \req{Weylcase}, the tangents to the ergosurface locii in the $(z,\rho)$ half-plane always have slopes $\pm \sqrt{2}$ at the merger point. An example of this type is Black Saturn. A Weyl solution with two unbalanced Myers-Perry black holes realizes the case of $\theta_\rom{m} = \frac{\pi}{2}$.

With some of the $U(1)$'s of the Weyl solutions replaced by spherical symmetry we can also derive a general result. If  the symmetries are $U(1)^k \times SO(d-k-2)$, then the result for the merger angle, again assuming singly spinning solutions only, is\footnote{Strictly speaking, one only needs the spherical symmetry to hold in a limiting sense at the merger point to obtain this result.}
\bea
  \label{dk2}
    \theta_\rom{m} = 2 \, \text{arccot}\sqrt{d-k-2} 
   ~~~~~~~\text{or}~~~~~~~~
    \theta_\rom{m} = \frac{\pi}{2} \, .
\eea 
In the first case the tangents to the ergosurface locii have slopes $\pm\sqrt{d-k-2}$ at the merger point. 

The curious result that the merger angle does not depend at all on the details of the solution but only on the dimensionality of spacetime and its symmetries is reminiscent of a similar result in a much simpler setting, namely mergers of equipotentials in Newtonian gravity! In the case of collinear sources, the tangents of the equipotentials have slopes $\pm\sqrt{d-1}$ and so the merger angle is $2\,\text{arccot}\sqrt{d-1}$, where $d+1$ is the spacetime dimension. This is simply a consequence of  Laplace's equation for a configuration with $SO(d-1)$ symmetry. 

We will show that the merger angle results quoted above for ergoregions is also a consequence of Laplace's equation. What happens is that at the point of merger, the relevant component of Einstein's equation simply reduces to Laplace's equation, and the conditions for an ergoregion merger point are completely equivalent to the conditions for a merger point of equipotential surfaces in Newtonian gravity. 

These statements are only true when the solution is singly spinning. When there is rotation in more than one plane, Einstein's equation at the merger point becomes a Poisson equation, since the interaction between multiple spins gives a source term. Universality of the merger angle is then lost. Explicit examples on doubly-spinning solutions indeed show that the merger angle can take a continuum of values as the parameters of the merger point solution are varied. 

The two cases of universal merger angles in both \req{Weylcase} and \req{dk2} appear to correspond to different co-dimensions $\delta$ of the merger surface: if the merger is extended along $p$ spatial directions, then $\delta=d-p$ and $\theta_\rom{m} = 2 \, \text{arccot}\sqrt{\delta-1}$. When the merger occurs at $\rho>0$ in the Weyl solutions it generically corresponds to a non-degenerate point for the orbits of the $U(1)$ rotational Killing vectors, and the merger surface therefore extends along all $p=d-2$ spatial Killing directions; thus $\delta=2$ and $\theta_\rom{m} = 2 \, \text{arccot}\sqrt{2-1} = \pi/2$. On the other hand, in a non-singular Weyl solution, a merger point  at $\rho=0$ corresponds to the fixed point of exactly one of the rotational isometries. The merger surface extends therefore only along $p=d-3$ spatial directions and hence $\delta=3$, so that $\theta_\rom{m} = 2 \, \text{arccot}\sqrt{2}$. The relationship between co-dimension and merger angle, $\theta_\rom{m} = 2 \, \text{arccot}\sqrt{\delta-1}$, yields an appealing formulation which precisely imitates that of Newtonian equipotentials.

The paper is structured as follows. To gain intuition for how the field equations imply universality of the merger angle, we analyze Newtonian equipotentials in \sec{laplaceprops}.  In \sec{proof}, we present the general analysis of ergosurface mergers in vacuum general relativity.  We first consider the generalized Weyl solutions and then  the case where some of the $U(1)$ symmetry is replaced by spherical symmetry. 
In \sec{univex} we demonstrate that our proof of universality covers non-trivial physical examples of interest.  We first analyze the Black Saturn family of solutions. Apart from confirming merger universality, we discuss under which circumstances mergers can happen. We then turn to the double Myers-Perry Weyl solution, which includes examples of both values \req{Weylcase} of the merger angle. Requiring rotation only in one plane is crucial to show universality, and to illustrate this point, we discuss  in \sec{nonunivex} two configurations where the merger angle is not universal. The orthogonal bi-ring solution is one example, the doubly-spinning black ring is the other.  The latter is interesting in its own right, since it illustrates that even a single regular black object can allow a topological transition of its ergosurface --- i.e.\ a self-merger.  In \sec{discuss} we summarize our results and discuss their implications.  Four short Appendices collect technical results relevant to the main text.

%____________________________________________
\section{Universality in Laplace equation }
\label{laplaceprops}
%____________________________________________

Let us briefly illustrate the salient properties of the Laplace equation,
\begin{equation}
\nabla^2 \Phi = 0 \ ,
\label{laplace}
\end{equation}	
first by considering the solutions exhibiting merger of level surfaces of the scalar field $\Phi$, and then by analyzing the equation itself.
To make contact with a specific familiar setup, we can consider the Newtonian potential $\Phi$ for a given source in $d$-dimensional space.  Recall that the potential for a point particle of mass $m$ in $d$ spatial dimensions is given by $\Phi(r) = - {m \over r^{d-2}}$, where $r$ is the distance from the source; for more general sources, we integrate this contribution over the entire source.

%~~~~~~~~~~~~~~~~~~~~~~~~~~~~~~~~~~~~~~~~~~~~~~~
\paragraph{Collinear point sources:}
%~~~~~~~~~~~~~~~~~~~~~~~~~~~~~~~~~~~~~~~~~~~~~~~

\begin{figure}[t]
\begin{center}
\includegraphics[width=7in]{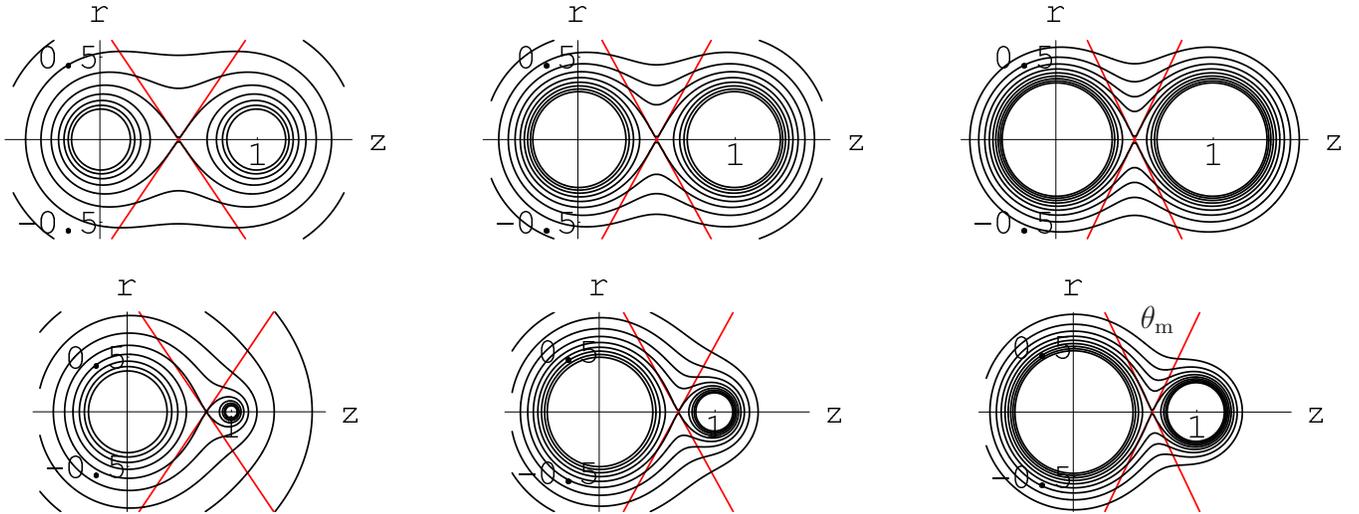}
\begin{picture}(0,0)(0,0)
\put(66,30){$\theta_\rom{m}$}
\end{picture}
\caption{Newtonian equipotentials (equally-spaced) for two point masses in various dimensions:  The first row has equal masses $m_1=m_2 = 1$, whereas the second row has differing masses, $m_1=10, m_2=1$.  The columns compare dimensionality; from left to right, the spatial dimension $d=3,4, 5$.  Superposed are the tangent lines at the merger.}
\label{nppts}
\end{center}
\end{figure}

First, as a warm-up, consider the Newtonian potential of two point particles of arbitrary masses  $m_1$ and $m_2$, separated by a distance which we will normalize to unity for convenience.  We will use cylindrical coordinates $(z,r,\Om_{d-2})$, so as to align the point particles along the $z$ axis and let $r$ denote the distance in the transverse directions.  The Newtonian potential in $d$ spatial dimensions is then given by 
\begin{equation}
\Phi(z,r)= -{m_1 \over (r^2 + z^2)^{d-2 \over 2}} 
  - {m_2 \over (r^2 + (z-1)^2)^{d-2 \over 2}} \, .
\label{nptwopts}
\end{equation}	
A selection of equipotentials, given by $\Phi(z,r)=const$ surfaces, are plotted on the $(z,r)$ plane in \fig{nppts}.  
% Figure 
We consider the equal mass case, $m_1=m_2$, in dimensions $d=3,4,5$ (first row of \fig{nppts}) as well as the case of $m_1=10\, m_2$ (second row). In each case there exists a value of the potential such that the corresponding equipotential surfaces of the point mass exactly touch each other at one point in the $(z,r)$-plane. We are interested in the angle between the tangents at the merger point.

We define the \emph{merger angle} $\theta_\rom{m}$ as indicated in \fig{fig:topch}: it is the angle between the tangents at the merger point, chosen in the region of the $(z,r)$-coordinate plane which does not contain the two sources of the gravitational field. In \fig{nppts}, the last plot indicates $\theta_\rom{m}$. Thus, if the slopes are $\pm \sigma$, then
\bea
 \theta_\rom{m} = 2 \, \rom{arccot}\, {\slope} ~~~\text{radians}\,.
\eea

Fig.~\ref{nppts} illustrates --- and we prove in the following ---
that the slopes of the tangents at the merger point depend only on the number of space dimensions $d$, but not on the masses or the separation. Thus, in a given dimension $d$, the merger angle $\theta_\rom{m}$ is universal. As $d$ increases, the merger angle becomes smaller. The faster falloff of
the Newtonian potential in higher dimensions allows the equipotential
surfaces of point particles to remain approximately spherical closer to the merger point and this results in steeper tangents (i.e.~larger slopes) at the merger point. Hence the smaller merger angles.

The merger angle can be found by expanding $\Phi(z,r)$ around the merger point $r=0,z=z_m$.  
At this point $\p_z \Phi = 0$ and $\p_r \Phi = 0$. Also, $\p_{r}\p_{z} \Phi = 0$, since the potential of collinear point sources is symmetric in $r \leftrightarrow -r$. Thus  to second order near the merger point we have 
\begin{equation}
\Phi(z,r) = \Phi(z_m,0) 
 + \half \, r^2 \, \p_r^2 \Phi(z_m,0) + \half \, (z-z_m)^2 \, \p_z^2 \Phi(z_m,0) 
 + \cdots 
\label{nptwoptexp}
\end{equation}	
The value of the potential at the merger point is $\Phi(z_m,0)$, and solving $\Phi(z,r) = \Phi(z_m,0)$ then gives 
\begin{equation}
r(z) = \pm \slope \, (z-z_m) 
+ \CO\( (z-z_m)^2 \) \, ,
\label{npequiprel}
\end{equation}	
where 
\begin{equation}
\slope =  \sqrt{-{\p_z^2 \Phi(z_m,0) \over \p_r^2 \Phi(z_m,0)}} \ .
\label{npequipslope}
\end{equation}	
Thus, in the $(z,r)$-plane, $\slope$ is the slope of the tangents to the equipotentials at the merger point, as plotted in \fig{nppts}.   Correspondingly, the angle between the equipotentials, determined by the slope $\slope$, is $\theta_\rom{m} = 2 \, \rom{arccot}\, {\slope}$ radians. 

Evaluating \req{npequipslope} for the two-particle potential \req{nptwopts} explicitly, we find that independently of the mass distribution, the merger slope, and hence $\theta_\rom{m}$, is universal and depends only on the number of space dimensions $d$:\begin{equation}
\slope = \sqrt{d-1} \, ,\hspace{1cm}
\theta_\rom{m} = 2 \, \rom{arccot}\, {\sqrt{d-1}} \, .
\label{slope2pt}
\end{equation}	
It is easy to verify that this result extends to multiple collinear point sources: each equipotential merger angle is independent of the mass distribution, with the slope given by \req{slope2pt}.  We show next that the universal behavior follows directly from the Laplace equation. 

%~~~~~~~~~~~~~~~~~~~~~~~~~~~~~~~~~~~~~~~~~~~~~~~
\paragraph{Sources with $d$-dimensional axial symmetry:}
%~~~~~~~~~~~~~~~~~~~~~~~~~~~~~~~~~~~~~~~~~~~~~~~

Consider in $d$ space dimensions sources arranged along a line such that transverse to the line the configuration has spherical symmetry $SO(d-1)$; this is what we mean by $d$-dimensional axial symmetry. 
%An example is $n \ge 2$ collinear point sources.

Let $z$ denote the coordinate along the axis singled out by the sources. The system is governed by the Laplace equation on flat $\mathbb{R}^{d}$ with a metric written in cylindrical coordinates as
\begin{equation}
ds^2 = dz^2 + dr^2 + r^2 \, d\Omega_{d-2}^2 \ . 
\label{flatmet}
\end{equation}	
By symmetry, the merger of equipotential surfaces must take place on the $z$-axis, i.e.~at points with $r=0$.
We focus on the near-merger region, that is, we consider a solution $\Phi$ which is constant on a cone:  $\pm r = \sigma \, (z- z_m)$ for some constants $\sigma$ and $z_m$. By shifting the $z$-coordinate we can take $z_m = 0$ without loss of generality. Also, by subtracting a constant, we can assume that $\Phi$ vanishes at the merger point. The potential must be smooth, so it takes the form
\begin{equation}
\Phi(z,r) = (r^2 - \sigma^2 \, z^2) \, f(z,r)  \ ,
\label{phismooth}
\end{equation}	
for some smooth function $f$.  
Substituting \req{phismooth} into the Laplace equation, we obtain
\bea 
\nonumber
0~=~\nabla^2 \Phi &=& 
2 \, [(d-1)- \sigma^2 ] \, f - 4 \,\sigma^2\, z \, \p_z f \\
&&+ \frac{1}{r}  \(  (d+2) \, r^2 - (d-2)\, \sigma^2\, z^2 \) \, \p_r f
+ (r^2 - \sigma^2 \, z^2) \, (\p_r^2 + \p_z^2 ) f \, .
\eea
Evaluating this at the point $z=r=0$, we find that only the first term survives; so in order to solve Laplace equation, we must have
$\slope = \sqrt{d-1}$.

The merger locus is defined by $\Phi(z,r)=0$. In the $(z,r)$ plane, the tangents of the locus at the merger point have slopes $\pm\sigma$. The calculation shows that $\sigma$ only depends on the dimension $d$. This makes universality manifest: the merger angle $\theta_\rom{m}$ does not depend on sources, as long as $\Phi$ remains a smooth function of $r$ and $z$ only.

% ------
\paragraph{Sources with translational invariance:}
% ----

Two parallel infinite uniform line sources are not covered by the case studied above. Instead the symmetry is $SO(d-2) \times \mathbb{R}$ (or $SO(d-2) \times U(1)$ if the direction along the sources is compactified to a circle). Consider more generally sources with symmetry $SO(m) \times U(1)^k $ in $d=m+k+1$ space dimensions. We write the metric on $\mathbb{R}^{d-k} \times T^k$ as
\begin{equation}
ds^2 = dz^2 + dr^2 + \delta_{ij} \,dy^i\,dy^j+ r^2 \, d\Omega_{m-1}^2 \, , 
\label{flatmet}
\end{equation}	
with the coordinates $y^i$ parameterizing the $k$ $U(1)$ symmetry directions; we may as well consider these to be non-compact. 

The potential depends as above only on $z$ and $r$, and the analysis of Laplace's equation proceeds as above. The result is that the slopes 
at a merger point and the merger angle are 
\begin{equation}
\slope = \sqrt{m} = \sqrt{d-k-1} \, ,\hspace{1cm}
\theta_\rom{m} = 2 \, \rom{arccot}\,{\sqrt{d-k-1}} \, .
\end{equation}	
Clearly, the toroidal directions play a passive role in the merger. Dimensional reduction gives the case of axial symmetry in $d-k$ dimensions.

% ------
\paragraph{Ring around central point source:}
% ----

Let us consider another example whose structure is very close to what we will encounter later when we study ergosurfaces.
Place a point source of mass $m_1$ at the center of a uniform circular ring source of mass density $m_2/(2\pi R)$, where $R$ is the radius of the ring. In $d$-dimensions the configuration has a $U(1)$ rotational symmetry in the plane of the ring, and transverse to this plane spherical symmetry $SO(d-2)$. By symmetry, the merger of equipotential surfaces occurs in the plane of the ring, along a circle concentric with the ring source. An explicit calculation shows that the slope of the tangents, at any point on the ``merger circle'', is independent of $m_1$, $m_2$, and $R$. We find, as the reader may already have guessed, simply
\begin{equation}
\slope = \sqrt{d-2}\, ,\hspace{1cm} 
\theta_\rom{m} =2 \, \rom{arccot}\,{\sqrt{d-2}}\ . 
\end{equation}	

This configuration can be thought of as a Newtonian Saturn system (without rotation). The full $4+1$-dimensional GR solution for Black Saturn, with the ring balanced by rotation, is one of our primary examples of ergoregion mergers. As we will show, the merger angle between ergosurfaces for the Black Saturn is $\sqrt{d-2} = \sqrt{2}$ and $\theta_\rom{m} \approx 70.53^\circ$.

%~~~~~~~~~~~~~~~~~~~~~~~~~~~~~~~~~~~~~~~~~~~~~~~
\paragraph{Non-axisymmetric sources:}
%~~~~~~~~~~~~~~~~~~~~~~~~~~~~~~~~~~~~~~~~~~~~~~~

Before turning to general relativity, let us emphasize the necessity of symmetry in our argument.   It is easy to demonstrate that universality of merger angles cannot hold in full generality. For instance, equipotential surfaces of non-collinear point sources in $d$ dimensions do not have universal mergers. 

A simple example illustrates the point. Consider in $d$-dimensions two parallel \emph{finite} uniform line distributions, each of length $L$. In the limit $L \to 0$, the rods become point sources and it follows from our calculation above that the tangents at the merger point of the equipotential surfaces have slopes $\slope=\sqrt{d-1}$. However, the limit $L \to \infty$ gives two parallel infinite line sources, and as we have shown, the merger slopes are then $\slope=\sqrt{d-2}$.
The finite length rod configurations interpolate between the extreme cases of $L=0$ and $L = \infty$, and the merger slopes are therefore be expected to vary continuously between $\sqrt{d-1}$ and $\sqrt{d-2}$ as $L$ increases. Thus the merger slopes, and hence the merger angles, for finite $L$ cannot be universal, but must depend on the details of the sources.

%~~~~~~~~~~~~~~~~~~~~~~~~~~~~~~~~~~~~~~~~~~~~~~~
\paragraph{Curved space:}
%~~~~~~~~~~~~~~~~~~~~~~~~~~~~~~~~~~~~~~~~~~~~~~~

The analyses presented in this section generalize to Laplace's equation in curved space. Using Riemann normal coordinates at the merger point, it is easy to see that only the local flat metric enters.  This observation will be borne out in the following section, where we extract universality directly from the Einstein's equation by showing that the relevant component reduces to the Laplace equation at points where ergosurfaces merge.

%____________________________________________
\section{General analysis of ergosurface merger}
\label{proof}
%____________________________________________
Having examined the constraints on mergers of Newtonian equipotentials we now turn to the problem at hand: mergers of ergosurfaces in general relativity. 
We show that --- with appropriate assumptions detailed in the following --- the relevant component of Einstein's equation reduces to the Laplace equation at the merger point. Following the basic strategy of \sec{laplaceprops}, we use this to prove that the ergosurfaces merge at a universal angle for solutions with angular momentum only in a single plane. 

We will focus on stationary spacetimes in $D=d+1$ spacetime dimensions. 
In \sec{weylergo} we consider generalized Weyl solutions, i.e.\ spacetimes possessing $d-1$
commuting 
Killing vector fields. All examples discussed
in Sections \ref{univex} and \ref{nonunivex}
fall in this class. 
In \sec{hdmergers}, we assume the existence of 2 commuting Killing vectors, one of which is the stationary time Killing vector and the other one a rotational symmetry. 
In addition we require spherical symmetry $SO(d-2)$, or a combination of spherical and toroidal symmetry. (We will see that one only needs the spherical symmetry in a limiting sense at the merger point, but it appears unlikely that this symmetry will be present if it is not present in the full spacetime.)

%____________________________________________
\subsection{Ergosurface mergers in generalized Weyl solutions}
\label{weylergo}
%____________________________________________

Consider a $d+1$-dimensional generalized Weyl metric of the form
\begin{equation}
  ds^2=G_{ab}\,dx^a dx^b+e^{2\nu}\left(d\rho^2+dz^2\right)\, .  
\label{Wmetric} 
\end{equation}
It is assumed that $\nu$ and the matrix  $G$ depend only on $\rho$ and $z$ and that  $\det G= - \rho^2$. $G$ encapsulates the metric functions along the $d-1$ Killing directions  
$\frac{\partial}{\partial x^a}$ of the spacetime. Einstein's equation  can then be written compactly as
\bea
  \label{Geq}
  &&G'' + \ddot{G}+\frac{1}{\rho}\, G' 
  ~=~G' \, G^{-1} \, G' + \dot{G} \, G^{-1} \, \dot{G} \, , \\[2mm]
  &&\nu' ~=~\frac{1}{2}
  \bigg[-\frac{1}{\rho}+\frac{\rho}{4}\,
  \Tr\left( (G' \, G^{-1})^2-(\dot{G} \, G^{-1} )^2\right)\bigg]\,,
  \hspace{5mm}
  \dot{\nu}~=~\frac{\rho}{4}\,\Tr\left(G' G^{-1} \dot{G} G^{-1}\right)\, ,
  ~~~~~~~
\eea
where for brevity $G' = \partial_\rho G$, $\dot{G} = \partial_z G$, $\nu'=\partial_\rho \nu$ and $\dot\nu=\partial_z\nu$.

Let us assume that the stationary metric \reef{Wmetric} describes an asymptotically flat spacetime\footnote{Our results will also apply to asymptotically Kaluza-Klein spacetimes.} and that the stationary time coordinate $x^0=t$ is canonically normalized at infinity. Thus we are assuming that asymptotically $G_{tt} \to -1$ and $G_{ta} \to 0$ for $a\ne t$.  The ergosurfaces, if present, are characterized by $G_{tt} = 0$. In general, this determines implicitly the coordinates $(z,\rho(z))$ of the ergosurface, and we will refer to the set of points $(z,\rho(z))$  as the \emph{ergosurface locus} in the $(z,\rho)$ half-plane.

If the spacetime has two disconnected ergoregions, then there will be two disjoint ergosurface locii in the $(z,\rho)$ half-plane. If, as parameters in a family of solutions are changed, the locii join, then there will be a special intermediate configuration at which the ergosurfaces intersect; this is the merger point. (This is completely analogous to the merger of equipotentials, but of course with a different physical interpretation.)
Locally, near a merger point, the locii can be described as the topology change sketched in \fig{fig:topch}. A merger point $(z_0,\rho_0)$ is therefore characterized by  
\bea
  \label{eqn:Econd}
  G_{tt}(z_0,\rho_0) = 0\, ,~~~~
  G_{tt}'(z_0,\rho_0) = 0\, ,~~~~
  \dot{G}_{tt}(z_0,\rho_0) = 0\, .
\eea

Near the merger point we expand $G_{tt}$ as
\bea
  G_{tt}(z,\rho) ~=~ \frac{1}{2}\, a\,  (\rho-\rho_0)^2 
  + \frac{1}{2}\, b  \,(z-z_0)^2 
  + c\, (\rho-\rho_0)(z-z_0) +\CO\left(\epsilon^3\right) \, ,
  \label{eqn:Gmerger}
\eea
where $a=G_{tt}''(z_0,\rho_0)$, $b=\ddot{G}_{tt}(z_0,\rho_0)$ and 
$c=\dot{G}_{tt}'(z_0,\rho_0)$.  We will not need the higher order corrections 
$\CO\left(\epsilon^3\right) = \CO\left((z-z_0)^3,(\rho-\rho_0)^3\right)$. 

Now solving $G_{tt}=0$ gives
\bea
  \label{slopes}
  \rho = \rho_0 + \alpha_\pm (z-z_0) + \dots \,  , \hspace{1.3cm}
  \alpha_\pm = \frac{1}{a} \Big( - c \pm \sqrt{c^2 - a b} \Big)\, ,
\eea
where ``\dots'' stand for quadratic and higher order terms.
It is clear from \reef{slopes} that near the merger point the ergosurface locii are approximately straight lines with slopes $\alpha_\pm$.

Note that when $c\neq 0$ the slopes $|\alpha_+|$ and $|\alpha_-|$ will not be equal in magnitude. To take this into account, one can rotate the coordinate system at the merger point such that in the new coordinates $\alpha_+' = - \alpha_-' > 0$. The merger angle $\theta_\rom{m}$, which we define as in \fig{fig:topch}, is of course invariant under this rotation. It is given by 
\bea
 \label{eqn:theta}
  \theta_\rom{m} = \pi - \arctan(\alpha_-') + \arctan(\alpha_+')
  = 2 \, \rom{arccot}\,
    \sqrt{\frac{-a-b-\sqrt{(a-b)^2+4c^2}}{a+b-\sqrt{(a-b)^2+4c^2}}}\, .
\eea
The argument of $\rom{arccot}$ in the second equality is simply the square-root of the ratio of eigenvalues of the second derivative matrix, the Hessian matrix of $G_{tt}$, evaluated at $(z_0,\rho_0)$. In examples where $c = 0$ we will discuss the slopes and the merger angle on equal footing, but when $c \ne 0$ we typically suppress the discussion of slopes unless referring explicitly to $\alpha_{\pm}$ as given in \req{slopes}.

We have not yet used  Einstein's equation, but will do so now to determine $a,b,c$ to the extent possible. The $tt$-component of Einstein's equation is
\begin{equation}
G_{tt}'' + \ddot{G}_{tt} +\frac{1}{\rho}\, G'_{tt}  = 
(G' \, G^{-1} \, G')_{tt} + (\dot{G} \, G^{-1} \, \dot{G})_{tt} \, .
\label{Gttfineq}
\end{equation}

Let us consider solutions with angular momentum only in a single plane.\footnote{Solutions with rotation in more than one plane are examined in \sec{nonunivex}.} Then the metric has just one off-diagonal component, say $G_{t\psi}$, and the terms on the rhs of \reef{Gttfineq} take the form
\bea
  \label{eqn:GGGtt}
  (G' \, G^{-1} \, G')_{tt} 
  =
  - \frac{1}{\rho^2}
  \left( G_{\psi\psi} (G_{tt}')^2 
  -2 G_{t\psi} G_{tt}' G_{t\psi}' 
  + G_{tt} (G_{t\psi}')^2 \right)
  \prod_{a\ne t, \psi} G_{aa} \, ,
\eea
and similarly for $ (\dot{G} \, G^{-1} \, \dot{G})_{tt}$.

In order to analyze \reef{Gttfineq} near the merger point $(z_0,\rho_0)$ 
we must consider the cases of $\rho_0 > 0$ and $\rho_0=0$ separately:

\begin{itemize}
\item \underline{$\rho_0 >0$}: 
When $\rho_0 > 0$ it is straightforward to see that in the limit $(z,\rho) \to (z_0,\rho_0)$, the conditions \reef{eqn:Econd} imply that the rhs of \reef{Gttfineq} vanishes. We are left with 
\begin{equation}
 G''_{tt} +\ddot{G}_{tt} = 0 \ ,
\label{offaxis}
\end{equation}
which gives $a=-b$. Inserting this into \reef{eqn:theta}, we find that
the merger angle is 
\be
\label{off}
\theta_\textrm{m}=\frac{\pi}{2}\, ,
\ee 
irrespective of the physical details of the solution.\footnote{Ref.~\cite{Chrusciel:2006se} derived the relation between the second derivatives of $G_{tt}$, $a=-b$,  for general Ernst solutions (or any Weyl solution) in four dimensions.}

\item \underline{$\rho_0 = 0$}: First note that Einstein's equation is symmetric under $\rho \leftrightarrow - \rho$. Hence any solution shares this symmetry and the expansion of $G_{tt}$ around $\rho=0$ cannot contain any odd powers of $\rho$. We conclude that $c=0$ in \reef{eqn:Gmerger}.

The limit $(z,\rho) \to (z_0,\rho_0)$ must be taken carefully since there are terms both on the lhs and rhs of \reef{Gttfineq} which naively behave as $0/0$. A detailed analysis of subleading terms in \reef{eqn:GGGtt} shows that the rhs of 
\reef{Gttfineq} vanishes. However, the $G_{tt}'/\rho$ term on the lhs does contribute, and $G_{tt}'' + \ddot{G}_{tt} +\frac{1}{\rho}\, G'_{tt} 
  \to a+ b+ a$. Thus  \reef{Gttfineq} gives
\bea
  \label{ab}
  b = -2a \, .
\eea 
This fixes the slopes at the merger and the merger angle to be
\bea
  \label{on}
  \alpha_\pm  = \pm \sqrt{2} \, ,\hspace{1.5cm}
  \theta_\rom{m} = 2\, \rom{arctcot}\,\sqrt{2} \sim 70.53^\circ \, ,
\eea
as can be seen from equations \reef{slopes} and \reef{eqn:theta}. 
\end{itemize}

To summarize, the merger angle $\theta_\rom{m}$ for ergosurface mergers in singly spinning generalized Weyl solutions is universal. It can only take two values, namely those in \req{off} and \req{on}, depending on whether the merger point is at $\rho > 0$ or $\rho=0$. As described in the Introduction, the difference between these two cases seem to be related to the co-dimension of the merger surface. We will present examples in \sec{univex}.

%________________________________________________
\subsection{Ergosurface mergers in stationary axisymmetric spacetimes}
\label{hdmergers}
%_______________________________________________

We now show that, under mild assumptions, ergosurface merger points are governed by the Laplace equation. This follows from an analysis of  Einstein's equation.
We start with stationary solutions with $\partial_t$ being the asymptotic time translation generator. A general metric of this form can be written as 
\begin{equation}
ds^2 = \G_{tt}\, dt^2 + 2\,  \omega_a \, dx^a\, dt + ds^2(\widetilde{\CB}) \, ,
\label{stymetgen}
\end{equation}	
where $\widetilde{\CB}$ is a co-dimension one surface in spacetime and $\omega = \omega_a \, dx^a$ is a one-form on $\widetilde{\CB}$. While we can analyse Einstein's equations for the metric \req{stymetgen}, we need to make the following assumptions about the 1-form $\omega$ defined on $\widetilde{\CB}$ to understand the behavior of ergosurfaces.  

%The simplest situation with any hope of universality is one where we have 
Let us first consider the simplest situation, with
singly spinning configurations, i.e.~a solution with non-vanishing angular momentum only in a single plane. We can then write $\omega = \omega_\psi\,d\psi$, where $\psi^a=(\partial_\psi)^a$ is the generator of the rotation.\footnote{For a black hole spacetime, the existence of the second Killing vector $\psi^a$ is guaranteed by the rigidity theorem, which was recently extended to $D>4$ dimensions in \cite{Hollands:2006rj,Moncrief:2008mr, Hollands:2008wn}.} We shall further assume that $(t,\psi)\leftrightarrow (-t,-\psi)$ is a symmetry of the spacetime, which is a reasonable physical requirement for any rotating body.\footnote{Technically, this is equivalent to demanding that $\psi^a$ be hypersurface orthogonal on $\widetilde{\CB}$. We require this in order to split the metric into a block-diagonal form as in \req{phystymet}.} So we focus on spacetimes whose metric takes the form
\begin{equation}
 ds^2= \G_{tt}\,dt^2+2\,\G_{t\psi}\,dt\,d\psi+\G_{\psi\psi}\,d\psi^2+ds^2(\CB)\,,
 \label{phystymet}
\end{equation}
where $\CB$, which we refer to as the base, is a Riemannian manifold with metric $h_{ab}$.

We need to determine the local geometry of the ergosurface merger, and to this end  we require information about  Einstein's equation. The equations of motion for metrics with two commuting Killing vectors, such as the ones we consider here, are derived in \App{geroch2}. (This is a straightforward generalization to $D=d+1$ dimensions of Geroch's work \cite{Geroch:1970nt,Geroch:1972yt}.) The result for Einstein's equation, expressed in terms of the metric components $\gamma_{ij}$, $i,j=t,\psi$, and the base metric $h_{ab}$ is given in
\req{eqn:scalar} and \req{eqn:ricci}. Here we only need the equation for $\G_{tt}$ which can be written
\begin{equation}
  D^aD_a\G_{tt}= \frac{\G_{tt}}{2\,\tau}\left[(D^a\G_{tt})(D_a\G_{\psi\psi})-2\,(D\G_{t\psi})^2\right]-\frac{1}{2\,\tau}\left[\G_{\psi\psi}(D\G_{tt})^2-2\,\G_{t\psi}(D^a\G_{t\psi})(D_a\G_{tt})\right]\,.
\label{eomF}
\end{equation}	
The $D_a$ are covariant derivatives with respect to the base metric $h_{ab}$ and $\tau = -{\rm det}\left(\G\right)$.

We will use the dynamical information contained in \req{eomF} to learn about the geometry of the ergosurface merger.  In fact, we will study this equation {\it at the merger point} $P$, to determine the angle between the merging surfaces. The rationale for this is that the merger angle is defined invariantly in terms of the inner product between the normals of the two components of the ergosurface as they intersect.
% This angle depends only on the local data at $P$ and so one can restrict attention to the tangent space of $\CB$ at $P$. 
To obtain this, we only need to evaluate \req{eomF} at $P$.

At the merger point we have $\G_{tt}|_P = D \G_{tt}|_P = 0$, which imply that 
$\G_{tt}$ has to be harmonic there
\begin{equation}
D^2 \G_{tt} \big|_P= 0 \ , 
\label{lapeqn}
\end{equation}	
since all terms in the rhs of \req{eomF} are explicitly proportional to $\G_{tt}$ or its first derivatives. Thus Einstein's equation for $\G_{tt}$ reduces to Laplace's equation at the merger point. Even though this equation only holds at one point, it is sufficient to determine the merger angle.  To see this, we proceed in close analogue with the Newtonian equipotentials in \sec{laplaceprops}.

We need to evaluate the Laplacian of $\G_{tt}$ at $P$, so we choose coordinates in a neighborhood of $P$ so that
\begin{equation}
h_{ab}\, dx^a \, dx^b \big|_P = dr_1^2 + dr_2^2 + r_2^2\, d\Omega_{d-3}^2 \ .
\label{tspace}
\end{equation}	
Since we have not yet assumed any symmetry on $\CB$, $\G_{tt}$ near $P$ can depend on all of these coordinates.
We now make the additional assumption that the ergosurface has $SO(d-2)$ symmetry near $P$, so that $\G_{tt}$ depends only on $r_1$ and $r_2$. (This will clearly be the case if $\CB$ itself has this symmetry.)
If the merger in the $(r_1,r_2)$ plane occurs at  $P= (r_{1*} , r_{2*})$ we can then expand 
\begin{equation}
\G_{tt}(r_1, r_2)  = \frac{a}{2}\, (r_1 - r_{1*})^2 +   c \, (r_1 - r_{1*})\,(r_2- r_{2*})+ \frac{b}{2}\, (r_2 - r_{2*})^2 \, ,
\label{fr1r2}
\end{equation}	
and \req{lapeqn} simply gives
\begin{equation}
\partial_1^2 \G_{tt} + \partial_2^2 \G_{tt} 
+ (d-3) \frac{1}{r_2}\, \partial_2 \G_{tt} = 0 \, .
\label{feqN}
\end{equation}	
It follows from \req{fr1r2} and \req{feqN} that 
\begin{eqnarray}
a &=& -b \qquad\qquad \qquad\qquad\qquad\qquad\,
 {\rm if} \;\;\; r_{2*} \neq 0 \, , \\
a &=& - (d-2)\, b ~\;\;\;\& \;\;\;~ c ~=~0 \qquad\;\;\; {\rm if} \;\;\; r_{2*} = 0 \, .
\label{}
\end{eqnarray}	
When $r_{2*} =0$, reflection symmetry in $r_2$ forces $c =0$. As in \sec{weylergo}, it follows that in the $(r_1, r_2)$-plane the merger angle is $\pi/2$ if $r_{2*} \neq 0$, and 
$2 \, \text{arccot}\sqrt{d-2} $
if $r_{2*} = 0$.
%the slopes of the tangents at the merger point are $\sqrt{d-2}$. 
Again, this is independent of all other details of the merger. 

If we had not assumed spherical symmetry $SO(d-2)$, but instead 
$SO(d-k-2) \times  U(1)^k$, then we would have found slopes $\sqrt{d-k-2}$ when $r_{2*}=0$. 
Comparing with \sec{laplaceprops} in which we analyzed Laplace's equation in $d$-dimensional flat space, here we have Laplace's equation in $d-1$ space dimensions, since the $\psi$-direction is treated separately. 

The Weyl solutions of \sec{weylergo} are simply the case of $k=d-4$. 
The sphere part of \req{tspace} is then just a circle; let it be generated by the Killing vector $\partial_\chi$. The $d-1$ commuting Killing vectors of the Weyl solution are then $\partial_t$, $\partial_\psi$, $\partial_\chi$ and the $k=d-4$ Killing vectors of the $k$ $U(1)$'s. Consistently, the slope of $\sqrt{2}$ was found in both analyses for the case of $r_{2*} = 0$ and the merger angle $\pi/2$ if $r_{2*} \neq 0$ (in Weyl coordinates, $\rho=0$ and $\rho>0$, resp.).

We noted earlier in \sec{weylergo} that when the solutions have angular momentum in multiple planes, it is harder to analyze the dynamical equation for $\G_{tt}$ at the merger point. The terms coming from the rotation couple to the equation determining $\G_{tt}$ in a non-trivial fashion.\footnote{To see this consider the simple extension of \req{eqn:scalar} with $\gamma$ elevated an $N\times N$ matrix encoding all the rotation terms. The non-trivial source for the Laplace equations comes from cross-terms $\gamma_{ti}$ on the rhs; it is the presence of this source which induces the non-universality. Essentially, multiple spins can conspire to  provide non-trivial source to the Laplace equation.} We know from explicit examples in five dimensional Weyl solutions that universality is lost when more than one angular momentum is turned on and it can be expected that this also holds for other solutions in higher dimensions.  

Finally, it is worth pointing out that the results obtained here will also be valid when a cosmological constant $\Lambda $ is included. The relevant effect of $\Lambda$ is an additional term proportional to $\Lambda \, \gamma_{tt}$ on the rhs of \req{eomF}; this follows from \req{eqn:scalar}. On the ergosurface, and hence in particular at the merger point, this term vanishes. 

%____________________________________________
\section{Examples of universal mergers}
\label{univex}
%____________________________________________

The two analyses of the previous section show --- for solutions of the vacuum Einstein's equation with spin in a single plane --- that \emph{if} ergosurfaces merge, then they merge with specific merger angles which depend only on the dimension of the spacetime and the symmetries of the solution.

The examples presented in this and the following section illustrate that there exist black hole solutions in which ergoregions merge and that the ergosurfaces can change topology as the solution parameters are varied. We consider in this section solutions with angular momentum in a single plane, and we show that they realize the specific merger angles found in the general analysis of \sec{proof}. 
Our primary example is the 4+1-dimensional Black Saturn solution, but we consider also the double Myers-Perry Weyl solution in 4+1-dimensions.

%____________________________________________
\subsection{Singly spinning Black Saturn ergosurface mergers}
\label{blacksaturn}
%____________________________________________

The Black Saturn solution \cite{Elvang:2007rd} consists of a black ring with horizon topology $\Sp^2 \times \Sp^1$ balanced by rotation around a 
%spherical $\mathbf{S}^3$ 
black hole with horizon topology $\Sp^3$. Each black object can carry angular momentum in the plane of the ring, and be co- or counter-rotating depending on the relative signs of their angular velocities. The system has 2-fold continuous non-uniqueness: fixing the ADM mass and angular momentum at asymptotic infinity, one can continuously vary the distribution of mass and angular momentum between the two objects. The requirement of balance fixes the radius of the ring. Thus, as parameters of the solution are varied, the proper distance between the black hole and the black ring will change, and 
they may be brought together close enough for the ergoregions to merge. The Saturn solution is exact, so the merger can by followed analytically and the solution parameters tuned precisely to the merger point. 

Ergoregion mergers take place only when the two ergoregions are co-rotating, so we consider the Saturn configurations in which the black hole and the black ring have angular velocities of the same sign. In fact, we focus primarily on the (simpler) subfamily of Saturn solutions in which the black hole has no intrinsic angular momentum, i.e.~vanishing Komar angular momentum $J_\rom{Komar}^\rom{BH}=0$. The black hole is nonetheless still rotating, it has non-vanishing angular velocity. This is due to rotational dragging by the surrounding rotating ring \cite{Elvang:2007rd}. Necessarily, the hole and the ring are then co-rotating, and there is 1-fold continuous non-uniqueness corresponding to distributing the total mass between the two objects. Near their horizons there are ergoregions: we are going to show here that when the two black objects are far apart the spacetime has two disjoint ergoregions, but when sufficiently close the ergoregions can merge.

All necessary properties needed to study the ergoregions of Black Saturn were presented in \cite{Elvang:2007rd}, and we will refer to this work for specific details while including here only a minimum of detail to keep the presentation clear. We proceed now to show that ergoregions can merge in the Saturn solution, we examine the conditions under which it happens, and finally we illustrate the merger location on a selection of branches of Saturn solutions in the ``phase diagram'' of 4+1-dimensional black holes.

%_________________________________________________
\subsubsection{Parameterization and constraints}
\label{SatParam}
%__________________________________________________

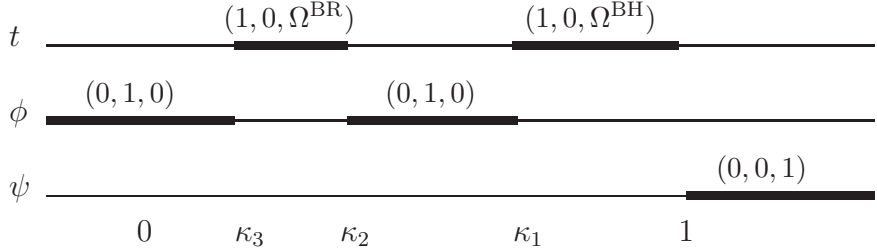
\begin{figure}[t]
\vspace{2cm}
\begin{center}
\begin{picture}(0,0)
\setlength{\unitlength}{1cm}
%lines and rods
\put(-5,1.5){\line(1,0){2.5}}
\put(-2.5,1.5){\linethickness{0.1cm}{\line(1,0){1.5}}}
\put(-1.,1.5){\line(1,0){7.}}
\put(1.2,1.5){\linethickness{0.1cm}{\line(1,0){2.2}}}
\put(1.,0.5){\line(1,0){5}}
\put(-5,0.5){\linethickness{0.1cm}{\line(1,0){2.5}}}
\put(-2.5,0.5){\line(1,0){1.75}}
\put(-1.0,0.5){\linethickness{0.1cm}{\line(1,0){2.25}}}
\put(-5,-0.5){\line(1,0){11}}
\put(3.5,-0.5){\linethickness{0.1cm}{\line(1,0){2.5}}}
%labels
\put(-5.5,1.5){$t$}
\put(-5.5,-0.5){$\psi$}
\put(-5.5,0.5){$\phi$}
%rod directions
\put(-2.65,1.75){\small{$(1,0,\Omega^\rom{BR})$}}
\put(1.35,1.75){\small{$(1,0,\Omega^\rom{BH})$}}
\put(-4.5,0.75){\small{$(0,1,0)$}}
\put(-0.5,0.75){\small{$(0,1,0)$}}
\put(3.9,-0.25){\small{$(0,0,1)$}}
%points
\put(-3.8,-1.1){$0$}
\put(-2.5,-1.1){$\ka_3$}
\put(-1.1,-1.1){$\ka_2$}
\put(1.2,-1.1){$\ka_1$}
\put(3.4,-1.1){$1$}
\end{picture}
\end{center}
\vspace{0.5cm}
\caption{\small{Rod structure of the Black Saturn.}}
\label{fig:saturn}
\end{figure}

The balanced Black Saturn solution is parameterized by an overall length scale $L$, and three dimensionless parameters $\kappa_{1,2,3}$. The solution is of the Weyl form \req{Wmetric} with Killing directions $x^a=t,\phi,\psi$, and it can be characterized in terms of its rod structure which is given in \fig{fig:saturn} (see \cite{Elvang:2007rd} for a details). The parameters $\kappa_{1,2,3}$ are directly related to the lengths of the rods, as shown in \fig{fig:saturn}, and must therefore satisfy the inequality 
\bea
  0 < \ka_3 \le \ka_2 < \ka_1 \le 1 \, .
\eea

The rods are located at $\rho=0$. The finite rods $z \in [\kappa_3,\kappa_2]$ and $[\kappa_1,1]$ are the locations of the black ring and black hole horizons, respectively.\footnote{We are working with dimensionless coordinates; dimensions are restored by multiplying $z$, $\rho$ and $\kappa_i$ by $L^{2}$.} Note that $\rho=0$ and $z \in [\kappa_2,\kappa_1]$ parameterize the part of the plane of the ring that lies between the ring and the hole horizons, while $\rho=0$  and $z \in (-\infty,\kappa_3]$ is the plane outside the ring. Finally $\rho=0$ and $z \in [1,\infty)$ is where the orbits of $\partial_\psi$ shrink to zero.

%figure

The general Saturn solution includes a fourth dimensionless parameter $\bar{c}_2$ which is fixed in terms of $\kappa_{1,2,3}$ as
\bea
  \bc_2 = \frac{1}{\ka_2}
  \left[
   \frac{\ka_1-\ka_2}
    { \sqrt{\ka_1 (1-\ka_2)(1-\ka_3)(\ka_1-\ka_3)} } - 1
  \right] \, . 
  \label{balance}
\eea
Equation \req{balance} is the balance condition which ensures that conical singularities are absent, so that the solution is regular on and outside the horizons.

The intrinsic angular momentum $J_\rom{Komar}^\rom{BH}$ vanishes if and only if $\bc_2 = 0$ \cite{Elvang:2007rd}. 
We comment briefly on $J_\rom{Komar}^\rom{BH}\ne 0$ at the end of the section, but will from now on  focus on the subfamily of Saturns with $J_\rom{Komar}^\rom{BH}=0$. The balance condition \req{balance} must then be solved with $\bar{c}_2=0$. It is convenient to solve for $\ka_3$; there are two solutions, but one is disgarded because it does not satisfy $\ka_3 \le \ka_2$. The valid solution is
\bea
  \label{k3star}
  \ka_3^* = \frac{1}{2} 
   \left(
     1+\ka_1 
     - \frac{\sqrt{\ka_1 (1-\ka_2) 
     \big[4 \ka_2 (2 \ka_1 - \ka_1^2 -\ka_2)
          -\ka_1 (1+\ka_1)^2 (1-\ka_2)\big]}}
     {\ka_1 (1-\ka_2)}
   \right) \, .
\eea
The expression under the square root is positive for
$0<\ka_2<\ka_1<1$, so the solution is real. It can also be seen that
$\ka_3^* \ge 0$, but the condition $\ka_3^* \le \ka_2$ requires that 
\bea
  \labell{k1min}
  \ka_1 \ge \ka_1^\rom{min} \equiv \frac{1}{2-\ka_2} > \ka_2 \, .
\eea
When the parameters $\ka_{1,2,3}$ of the solution are varied it must be done subject to the conditions \req{k3star} and \req{k1min}.

It is useful for the interpretation of the results to replace one of the $\ka_i$'s with the Komar mass of the black hole. In all applications here, we eliminate the overall scale $L$ of the solution by fixing the ADM mass of the system and work with dimensionless variables. The ratio of the black hole Komar mass to the total ADM mass is\footnote{This follows from (3.34) and (3.30) of \cite{Elvang:2007rd}.}
\bea
  \label{littlem}
  m \equiv \frac{M^\rom{BH}_\rom{Komar}}{M} = \frac{1-\ka_1}{1-\ka_1 +\ka_2} \, .
\eea
Note that $0 \le m < 1$. We can consider $m$ an estimate of how much of the total mass is located in the black hole. Note that the sum of the black hole and black ring Komar masses equals the ADM mass, since the solution solves Einstein's equation in vacuum and has no naked singularities.

Solving this for $\ka_1$ in \req{littlem} gives
\bea
  \labell{k1star}
  \ka_1^* = 1 - \frac{m}{m-1}\,  \ka_2 \, .
\eea
The condition $\ka_1^\rom{min}<\ka_1^*$ of \reef{k1min} now  requires for given $m$ that
\bea
  \ka_2 < \ka_2^\rom{max}(m)
  \equiv \frac{1+m-\sqrt{1-2 m + 5m^2}}{2 m} \, .\label{k2max}
\eea 

To summarize, for given relative mass $m$ of the Myers-Perry black
hole of the Saturn system with $J_\rom{Komar}^\rom{BH}=0$, we have one free parameter, $\ka_2$, which takes values 
\bea
  0 < \ka_2 < \ka_2^\rom{max}(m) \, ,
\eea
and $\ka_{1}$ and $\ka_{3}$ are fixed by \reef{k3star} and \reef{k1star}. Since we have fixed the total mass $M$ and the mass of the central black hole $m$, the free parameter corresponds to changing the ADM angular momentum. As this is done, the ring radius varies accordingly to maintain balance.

%______________________________________________________
\subsubsection{Mergers do happen!}
\label{satergplane}
%______________________________________________________

The location of the ergosurface(s) is found by solving the equation $G_{tt}(z,\rho)=0$. For Black Saturn $G_{tt}=-H_y(z,\rho)/H_x(z,\rho)$ is a rather involved function of $(z,\rho)$. Its specific form is given by equations (2.26)-(2.33) of \cite{Elvang:2007rd}.

In the limit $\rho \to 0$, the equation $G_{tt}(z,\rho)=0$ becomes more tractable. We take the limit subject to the condition that $z \in [\kappa_3,\kappa_2]$. This means that we restrict ourselves to the plane of the ring, between the ring and hole horizons. The plane of the ring is a plane of symmetry of the solution, so if there are two disjoint ergosurfaces, they must necessarily both intersect the plane between the black hole and the black ring. This means that $G_{tt}(z,0)=0$ must have two real roots $z_\pm \in [\kappa_3,\kappa_2]$. 
If there are no such solutions, then there cannot be two disjoint ergosurfaces, and instead the two rotating objects must be surrounded by a single ergoregion with a single component ergosurface which only intersects the plane of the ring outside the ring. Thus, when $G_{tt}(z,0)=0$ has a double root, $z_+=z_-$, this is exactly the merger point of interest. 

To be specific, when $J_\rom{Komar}^\rom{BH}=0$, the equation  $G_{tt}(z,0)=0$ gives the simple polynomial
\begin{equation}
z^2  - z \big[  \ka_1 + \ka_2 + \ka_3 - \ka_1 \, \ka_2 \, \ka_3) \big]
+ \ka_1 \, \ka_2 + \ka_2 \, \ka_3 + \ka_3 \, \ka_1 - 2\, \ka_1 \, \ka_2 \, \ka_3 
  ~ = ~ 0 \, .
\label{gttz}
\end{equation}
Reality of the roots requires
\begin{equation}
    f(\kappa_1,\kappa_2,\kappa_3) \;\equiv\;  \big[  \ka_1 + \ka_2 + \ka_3 - \ka_1 \, \ka_2 \, \ka_3) \big]^2
    - 4 \big[ \ka_1 \, \ka_2 + \ka_2 \, \ka_3 + \ka_3 \, \ka_1 
  -       2 \ka_1 \, \ka_2 \, \ka_3 \big] \;\ge\; 0 \, ,
 \label{disc}
\end{equation}
with equality corresponding to the merger point. 

To demonstrate that mergers do occur, one simply has to show that the function $f=f(\ka_i)$ changes sign when the parameters $\ka_i$
are varied subject to the conditions described in \sec{SatParam}. One easily finds that the ergosurfaces of the Saturn system do merge!

%figure
\begin{figure}[t!]
\begin{center}
\begin{tabular}{ccc}
\includegraphics[scale=0.62]{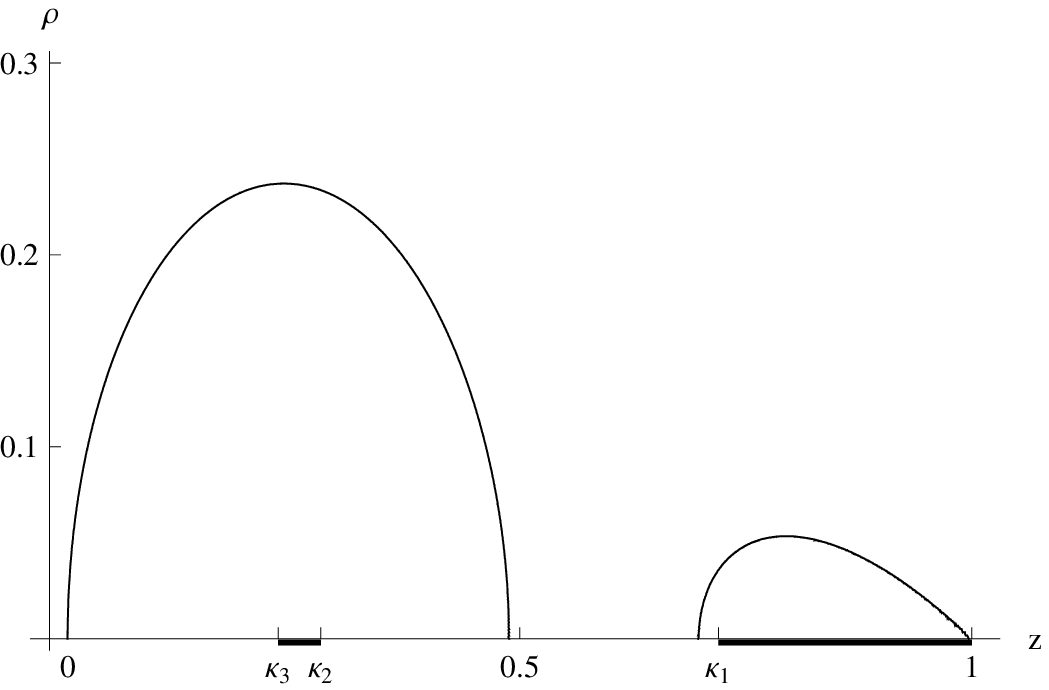}
&\hspace{0.3cm}&
\includegraphics[scale=0.62]{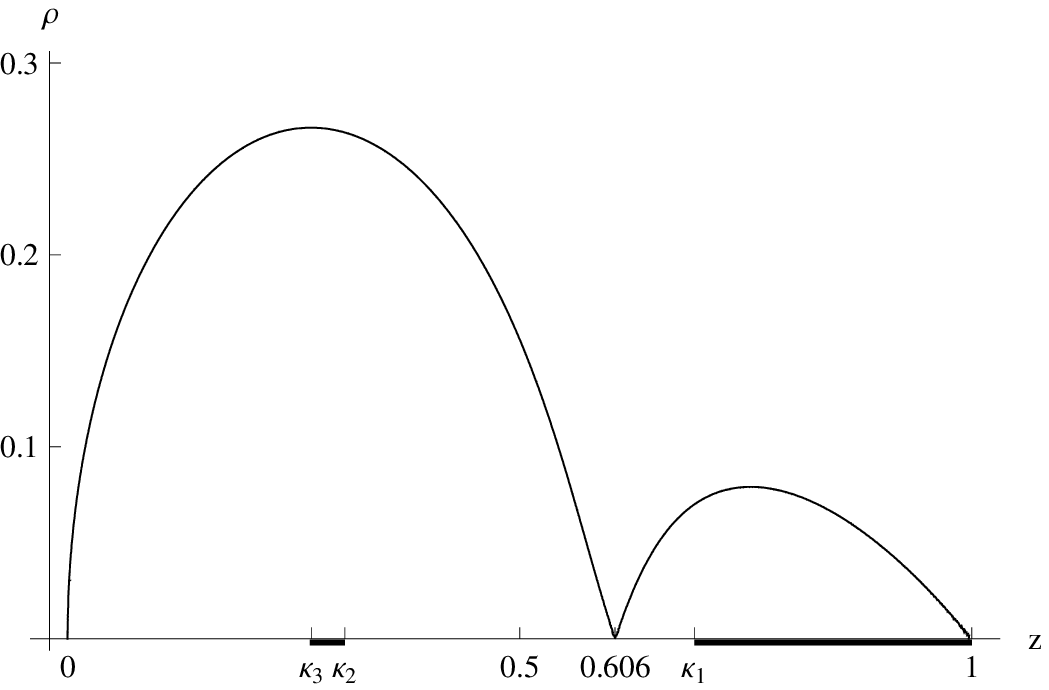} \\
\includegraphics[scale=0.62]{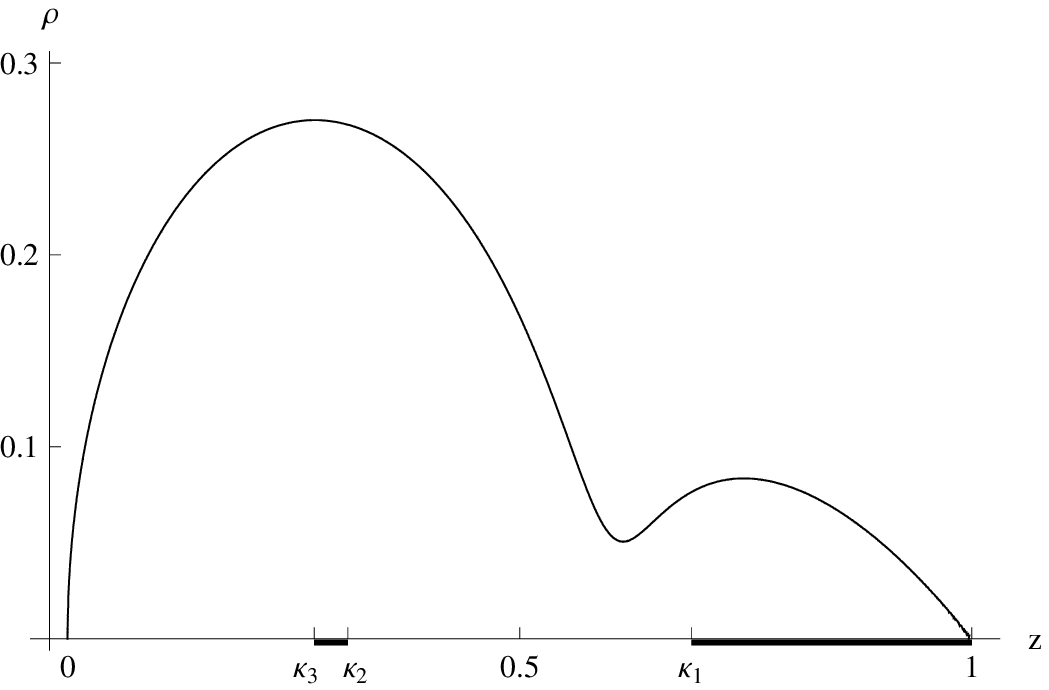}
&\hspace{0.3cm}&
\includegraphics[scale=0.62]{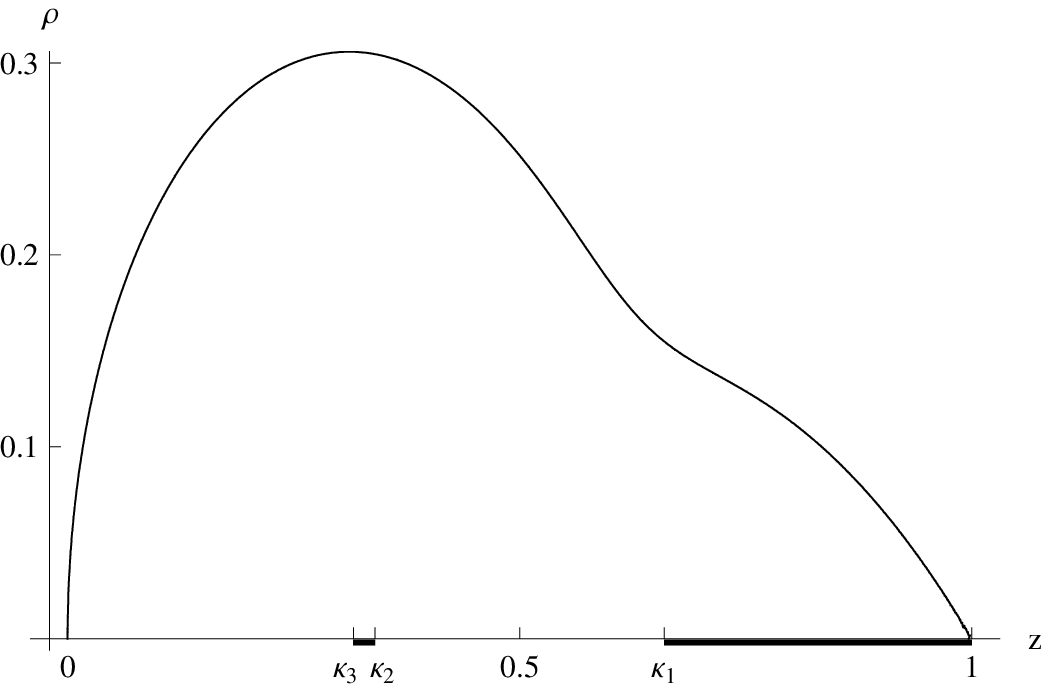}
\end{tabular}
\begin{picture}(0,0)
\setlength{\unitlength}{1cm}
%lines and rods
\put(-8.5,4){(a)}
\put(-1,4){(b)}
\put(-8.5,-0.6){(c)}
\put(-1,-0.6){(d)}
\end{picture}
\end{center}
\caption{Sequence of Black Saturn ergosurface locii for fixed 
$m_\textrm{BH}=0.5$ and $J_\rom{Komar}^\rom{BH}=0$. The black horizontal rods $[\kappa_3,\kappa_2]$ and $[\kappa_1,1]$ show the locations of the black ring and hole horizons, resp; compare with \fig{fig:saturn}.
(a) The black ring and black hole are widely separated. The ring has its own ergoregion with $\Sp^2\times \Sp^1$ ergosurface and the hole has a $\Sp^3$ ergosurface, which touches the $\Sp^3$ horizon at the pole at $z=1$. 
(b) Merger configuration ($(\kappa_2)_\textrm{merger}\approx 0.307$): the ergosurfaces are just touching. The slopes of the tangents at the merger are $\pm \sqrt{2}$. (c) There is a single ergosurface with topology $\Sp^3$ surrounding both the black ring and the black hole. The ergosurface still always touches the black hole horizon at the pole at  $z=1$. (d) The $\Sp^3$ ergosurface tends to become rounder as the ring and hole come closer.}
\label{fig:saturnmerger}
\end{figure}

It is illustrative to plot a sequence of ergosurface locii in the $(z,\rho)$-plane. For this we solve $G_{tt}(z,\rho)=0$ numerically.
In \fig{fig:saturnmerger} shows the result for fixed black hole mass $m_\textrm{BH}=0.5$. 
When the two black objects are far apart, as in \fig{fig:saturnmerger}(a), there are two disjoint ergosurfaces, and as they come closer they merge. \fig{fig:saturnmerger}(b) is precisely the configuration at the point of merger. In \fig{fig:saturnmerger}(c) and \fig{fig:saturnmerger}(d) the ergoregions have merged, and the ergosurface has changed topology from the disjoint set of $\Sp^2 \times \Sp^1$ and $\Sp^3$ to a single $\Sp^3$ surrounding both the hole and the ring --- it tends to become rounder as the black ring comes closer to the Myers-Perry black hole.\footnote{Note that for Weyl solutions, the topology change can be inferred directly from the rod diagrams such as \fig{fig:saturn}. In all plots in the $(z,\rho)$-plane we indicate the rod structure by labeling the end-points along the $z$-axis, see e.g.~\fig{fig:saturnmerger}. The ergosurface topology is obtained by looking at the fixed points of the rotational Killing fields. For instance, an ergosurface that intersects the $z$-axis on two different rods lying along the $\phi$-axis will have topology $\Sp^2 \times \Sp^1$. See \App{topology} for further details.} Note: the distance along the $z$-axis does not properly represent the distance between the two objects. Instead one should use the proper distance between the black hole and the black ring in the plane. This was done in \cite{Elvang:2007rd}, and those results confirm that the proper distance decreases monotonically along the sequence of plots shown in \fig{fig:saturnmerger}.

%___________________________________________
\subsubsection{Merger angle and slopes}
\label{Satangle}
%______________________________________________
Our general analysis of \sec{weylergo} for Weyl solutions tells us that the merger angle in the Saturn ergosurface mergers must be $\theta_\rom{m}\sim 70.53^\circ$ and that the slopes of the tangents at the merger point are  $\sigma = \pm \sqrt{2}$,
since this is a case where the merger point is at $\rho=0$. This is easily confirmed by an explicit calculation using the exact Saturn solution. We note also that the merger point extends along a circle in the plane of the ring. The plane is characterized as the location where the azimuthal circles of the $\mathbf{S}^2$ shrink to zero size. So the merger ``surface'' (spatially) 1-dimensional, and hence the co-dimension is $\delta=4-1=3$. The result $\sigma = \sqrt{2}$ matches the expectation $\sqrt{\delta-1}$.

%___________________________________________
\subsubsection{When do mergers happen? Critical BH mass}
\label{massangmerg}
%______________________________________________

When does the black ring get close enough to the black hole for the ergoregions to merge? Intuitively, if the black hole is too ``big'', the ring cannot rotate fast enough to balance itself very close to the black hole, and then the ergoregions may not merge. 
This is verified by the existence of a ``critical'' black hole mass $m_c$ which is an upper bound for Saturn mergers.

To determine $m_c$, we must solve $f=0$ with $f$ given by \reef{disc} with 
$\ka_1=\ka_1^*$ and $\ka_3=\ka_3^*$ from \req{k1star} and \req{k3star}. This gives an equation in $\ka_2$ and $m$. We note that the limit $\ka_2 \to 0$ corresponds to the large radius limit of the ring, whereas $\ka_2 \to \ka_2^\rom{max}$, with $\ka_2^\rom{max}$ of \req{k2max}, is the point of closest approach (minimal proper distance) between the black hole and the black ring. At the critical value $m_c$, the ergoregions would just touch at the points of closest approach. So to find $m_c$, we plug in $\ka_2 = \ka_2^\rom{max}$ into \reef{disc}. If the above substitutions are done so that surds are systematically eliminated from the equation, then one can solve for $m_c$ exactly. The result is
\bea
  m_c = \frac{20}{29} \approx  0.6897\dots
\eea

The corresponding critical values of the total (dimensionless) angular momentum and black hole Komar mass are\footnote{We have used \req{littlem} and eqs.~(3.30)-(3.31) and (4.1) of \cite{Elvang:2007rd}.}
\bea
  j_c^2=\frac{3^8}{(29)^{3}}
  \approx  0.269\dots \, ,
  \hspace{1cm}
  a_{\rom{H}\,c}^\rom{BH} = \frac{10^{5/2}}{29^{3/2}}
  \approx  2.02\dots
  \label{jcac}
\eea 
Mergers happen for $j_c < j < 1$ and $a_{\rom{H}}^\rom{BH} < a_{\rom{H}\,c}^\rom{BH}$.

Note that the dimensionless area and angular momenta are defined as
\begin{equation}
 j^2=\frac{27\pi}{32\,G_5}\,\frac{J^2}{M^3}\;,\qquad a_{\textrm{H}}^i =\frac{3}{16}\sqrt{\frac{3}{\pi}}\frac{{\cal A}_i}{(G_5 M)^{3/2}}\;,
\label{eqn:redpar}
\end{equation}
where the script $i$ labels either the black hole ($i=$ BH) or the black ring ($i=$ BR). The angular momentum $J$, areas ${\cal A}_i$, and mass $M$ are given in (3.31), (3.26)-(3.27), and (3.30) of \cite{Elvang:2007rd}.
The total area is $a_{\textrm{H}}^\rom{total}=a_{\textrm{H}}^\textrm{BR}+a_{\textrm{H}}^\textrm{BH}$.

%_________________________________________
\subsubsection{Phase diagram}
\label{saturnphys}
%__________________________________________

Consider the ``phase diagram'' showing total area, $a_{\textrm{H}}^\rom{total}$, versus total angular momentum squared, $j^2$, for fixed ADM mass $M$. We will illustrate here where on the Saturn ``phases'' branches the merger occurs.  We only consider Saturn configurations with $J_\rom{Komar}^\rom{BH}=0$. 

\begin{figure}[t]
 \begin{center}
\includegraphics[scale=1.25]{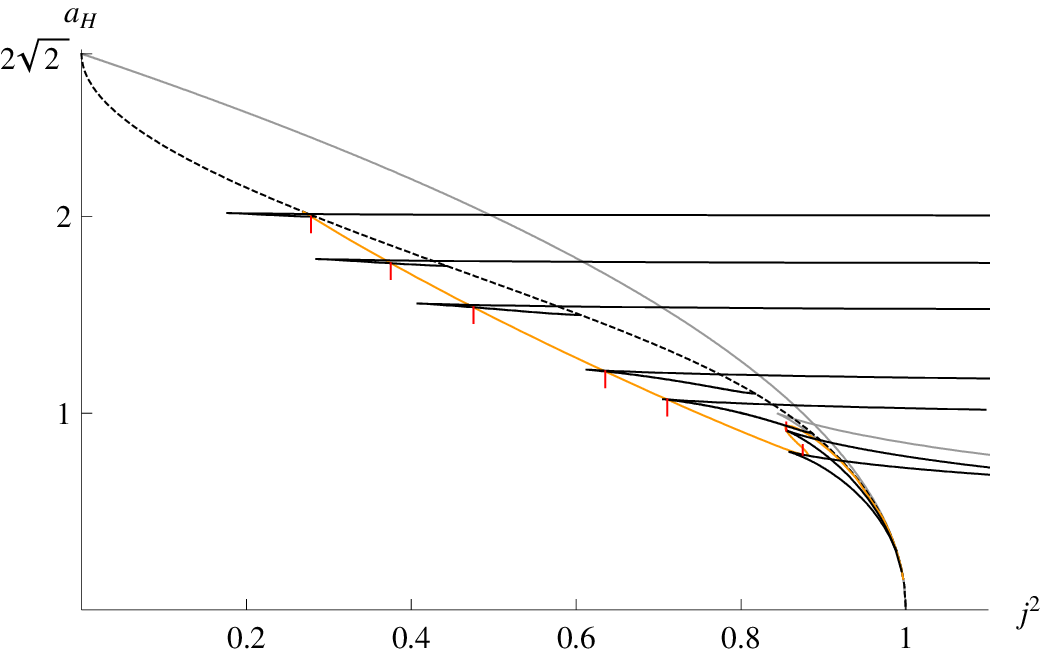}
\begin{picture}(0,0)
\setlength{\unitlength}{1cm}
\put(-9.7,7){\footnotesize MP}
\put(-1,2.67){\footnotesize BR}
\put(-0.35,2.35){\footnotesize $a_{\textrm{H}}^\textrm{BH} = 0.03$}
\put(-1.35,1.85){\footnotesize $a_{\textrm{H}}^\textrm{BH} = 0.3$}
\put(-0.35,3){\footnotesize $a_{\textrm{H}}^\textrm{BH} = 0.9$}
\put(-0.35,3.55){\footnotesize $a_{\textrm{H}}^\textrm{BH} = 1.1$}
\put(-0.35,4.35){\footnotesize $a_{\textrm{H}}^\textrm{BH} = 1.5$}
\put(-0.35,4.95){\footnotesize $a_{\textrm{H}}^\textrm{BH} = 1.75$}
\put(-0.35,5.55){\footnotesize $a_{\textrm{H}}^\textrm{BH} = 2$}
\end{picture}
\end{center}
\vspace{-4.5cm}
\hspace{3cm}
\includegraphics[scale=0.4]{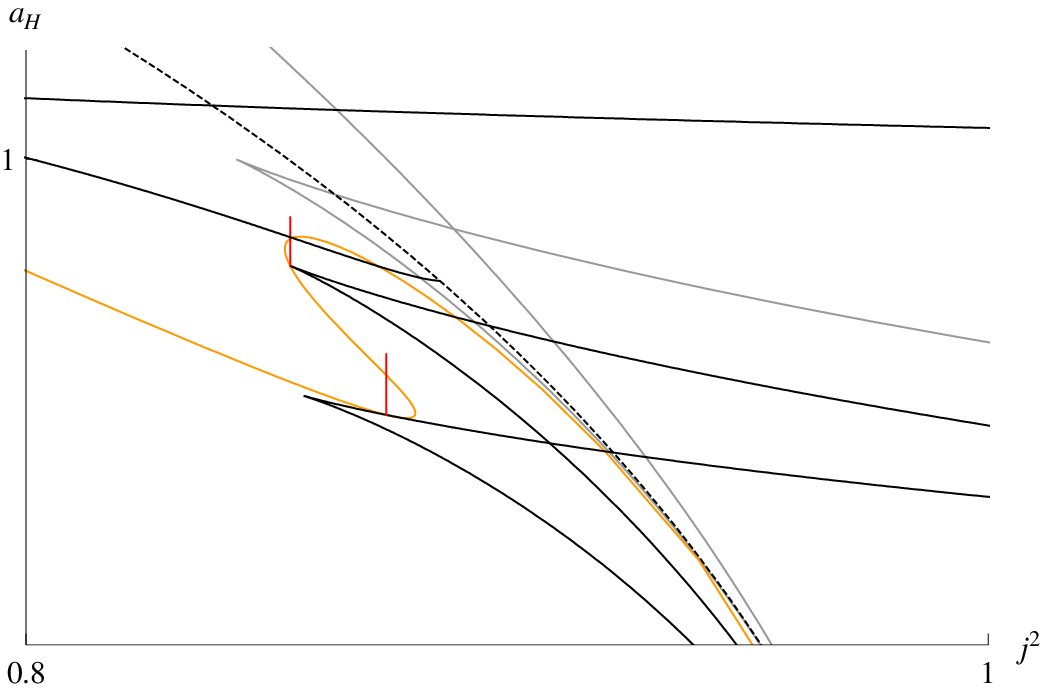}
\vspace{1.5cm}
\caption{Phase diagram with total horizon area $a_{\textrm{H}}^\textrm{total}$ versus total angular momentum squared $j^2$. The \emph{black} curves show branches of Saturn
configurations with fixed $a_{\textrm{H}}^\textrm{BH} = 0.03, 0.3, 0.9, 1.1, 1.5, 1.75, 2$. The \emph{red} indicators show where on each branch the ergoregion merger happens. A red indicator over the branch means the merger is on the upper thin ring branch, under means the merger is on the fat ring branch. The \emph{orange} curve is the curve of exact merger points.
The \emph{gray} curves show the branches of the singly spinning Myers-Perry  black hole and the single black ring. The \emph{dotted} curve is the end of the fat Saturn branches (corresponding to a Myers-Perry black hole surrounded by a nakedly singular black ring).}
\label{fig:atotvsjsq}
\end{figure}

If the area $a_{\textrm{H}}^\textrm{BH}$ of the black hole is held fixed,  then --- as the angular momentum of the ring is varied --- the Saturn configuration has a thin and a fat ring branch in the phase diagram. The thin and fat branches meet at a cusp. Several examples are shown in \fig{fig:atotvsjsq} (black curves; for reference, shown in gray are the phases of the Myers-Perry black hole as well as the black ring).\footnote{For more details on black ring thin and fat branches, see \cite{Elvang:2006dd, Emparan:2006mm, Emparan:2008eg}. The phases of Saturns shown here are of the same as those in Fig.~5 of \cite{Elvang:2007rd}.}
On the thin ring branches, the separation between the black hole and the black ring can become arbitrarily large as the angular momentum is increased. On the fat ring branch, the ring flattens out as $j$ increases towards a maximum at which the ring becomes singular. The dotted curve in \fig{fig:atotvsjsq} outlines these fat branch endpoints.

Without a central black hole, the ergosurface of the singly spinning black ring only self-intersects in the singular limit $j \to 1$, $a_\rom{H} \to 0$ where the fat ring branch ends. When a small black hole is present, i.e.~$a_\textrm{H}^\textrm{BH}$ is very small, the ergosurface merger will take place near the end of the fat branch of the Saturn phase. As $a_\textrm{H}^\textrm{BH}$ is taken to be larger, the merger point creeps up the fat ring branch towards the cusp where the thin and fat ring branches meet. The smallest fixed black hole area curve shown in \fig{fig:atotvsjsq} has $a_\textrm{H}^\textrm{BH}=0.03$, and here the merger point has already come up very close to the cusp, as shown by the red indicator line (see also insert in \fig{fig:atotvsjsq} which zooms in on this region). For the next value,  $a_\textrm{H}^\textrm{BH}=0.3$, the merger happens on the thin ring branch, but still close to the cusp. For all subsequent values shown, $a_\textrm{H}^\textrm{BH}=0.9, 1.1, 1.5, 1.75, 2$ the merger point has again come down on the fat ring branch. As the black hole area is increased towards the critical value 
$a_{\textrm{H}\,c}^\textrm{BH} \approx 2.02$ of \req{jcac}, the merger point again approaches the endpoint of the fat ring branch where it eventually disappears when $a_\textrm{H}^\textrm{BH}=a_{\textrm{H}\,c}^\textrm{BH}$.

%figure

%_________________________________________
\subsubsection{Saturn with $J_\rom{Komar}^\rom{BH} \ne 0$}
\label{genSat}

When $J_\rom{Komar}^\rom{BH} \ne 0$, merger of the ergoregions require that the black hole and black ring are co-rotating, i.e.~their angular velocities have the same sign. Such configurations exist and we have verified in examples that they exhibit mergers. The slopes of the tangents at the merger points are confirmed to be the expected $\pm \sqrt{2}$.

%____________________________________________
\subsection{Multi-ring configurations}
\label{ssdiring}
%____________________________________________

Saturn configurations with multiple rings can be constructed with the inverse scattering method, as can di-ring (or multi-ring) systems without the central black hole. According to our general analysis, these singly spinning configuration will also have universality when the ergoregions merge. The merger slope is $\pm\sqrt{2}$; we have confirmed this explicitly for the di-ring system which was previously constructed in \cite{Iguchi:2007is,Evslin:2007fv}. The physics of mergers in this case parallels that of Black Saturn. 

One can also consider the bi-ring system \cite{Izumi:2007qx,Elvang:2007hs} with spin in a single plane. This solution is not balanced.  However, one does have merger configurations, where merger occurs for $\rho_0 \neq 0$ and leads therefore to a merger angle of $\pi/2$ universally. We will return to a discussion of the bi-ring system with spins in both planes in \sec{s:birings}.

%____________________________________________
\subsection{Singly spinning double Myers-Perry configurations}
\label{ssdmp}
%____________________________________________

The ``true'' double Myers-Perry black hole solution is expected to have only one rotational $U(1)$ symmetry. This solution does not belong to the generalized Weyl class and has, as of now, not been constructed. Within the class of Weyl solutions, however, there are two distinct rod configurations that describe two static $\mathbf{S}^3$ black holes held apart by conical singularities. Rotation can be added using the inverse scattering method, and the result are two distinct Weyl-type double Myers-Perry black hole configurations.  One of these cases was recently studied in \cite{Herdeiro:2008en}. We consider the other case here, but focus on the subfamilies with angular momentum only in one plane of rotation. Balance can never be achieved in these configurations which therefore have conical singularities that keep the two black holes apart. This, however, does not interfere with our analysis of the ergoregion mergers.

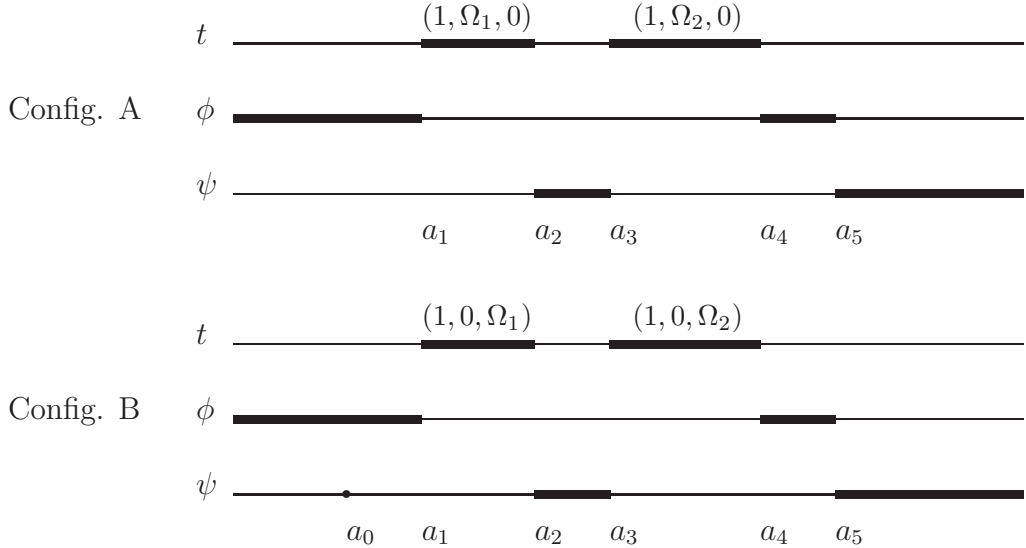
\begin{figure}[t]
\vspace{2cm}
\begin{center}
%first config
\begin{picture}(0,0)
\setlength{\unitlength}{1cm}
%lines and rods
\put(-5,1.5){\line(1,0){2.5}}
\put(-2.5,1.5){\linethickness{0.1cm}{\line(1,0){1.5}}}
\put(-1.,1.5){\line(1,0){1}}
\put(0,1.5){\linethickness{0.1cm}{\line(1,0){2}}}
\put(2,1.5){\line(1,0){3.5}}
\put(-5,0.5){\linethickness{0.1cm}{\line(1,0){2.5}}}
\put(-2.5,0.5){\line(1,0){4.5}}
\put(2.,0.5){\linethickness{0.1cm}{\line(1,0){1.0}}}
\put(3,0.5){\line(1,0){2.5}}
\put(-5,-0.5){\line(1,0){4}}
\put(-1.0,-0.5){\linethickness{0.1cm}{\line(1,0){1}}}
\put(0,-0.5){\line(1,0){3}}
\put(3,-0.5){\linethickness{0.1cm}{\line(1,0){2.5}}}
%labels
\put(-5.5,1.5){$t$}
\put(-5.5,-0.5){$\psi$}
\put(-5.5,0.5){$\phi$}
\put(-8.0, 0.5){\textrm{Config. A}}
%rod directions
\put(-2.5,1.75){{\small$(1,\Omega_1,0)$}}
\put(0.3,1.75){\small{$(1,\Omega_2,0)$}}
%points
\put(-2.5,-1.1){$a_1$}
\put(-1,-1.1){$a_2$}
\put(0,-1.1){$a_3$}
\put(2,-1.1){$a_4$}
\put(3,-1.1){$a_5$}
%second config
%lines and rods
\put(-5,-2.5){\line(1,0){2.5}}
\put(-2.5,-2.5){\linethickness{0.1cm}{\line(1,0){1.5}}}
\put(-1.,-2.5){\line(1,0){1}}
\put(0,-2.5){\linethickness{0.1cm}{\line(1,0){2}}}
\put(2,-2.5){\line(1,0){3.5}}
\put(-5,-3.5){\linethickness{0.1cm}{\line(1,0){2.5}}}
\put(-2.5,-3.5){\line(1,0){4.5}}
\put(2.,-3.5){\linethickness{0.1cm}{\line(1,0){1.0}}}
\put(3,-3.5){\line(1,0){2.5}}
\put(-5,-4.5){\line(1,0){4}}
\put(-3.5,-4.5){\circle*{0.1}}
\put(-1.0,-4.5){\linethickness{0.1cm}{\line(1,0){1}}}
\put(0,-4.5){\line(1,0){3}}
\put(3,-4.5){\linethickness{0.1cm}{\line(1,0){2.5}}}
%labels
\put(-5.5,-2.5){$t$}
\put(-5.5,-4.5){$\psi$}
\put(-5.5,-3.5){$\phi$}
\put(-8.0, -3.5){\textrm{Config. B}}
%rod directions
\put(-2.5,-2.25){{\small$(1,0,\Omega_1)$}}
\put(0.3,-2.25){\small{$(1,0,\Omega_2)$}}
%points
\put(-3.5,-5.1){$a_0$}
\put(-2.5,-5.1){$a_1$}
\put(-1,-5.1){$a_2$}
\put(0,-5.1){$a_3$}
\put(2,-5.1){$a_4$}
\put(3,-5.1){$a_5$}
\end{picture}
\end{center}
\vspace{4.5cm}
\caption{\small{Rod structure for the two singly spinning double Myers-Perry Weyl solutions.}}
\label{fig:ssdmp}
\end{figure}

The rod configurations for the double Myers-Perry Weyl solutions of our interest are given in \fig{fig:ssdmp}. The rod diagrams discriminates between the two rotational Killing vectors, $\partial_\phi$ and $\partial_\psi$, in that one of them shrinks to zero in a plane between the two black holes and the other one does not. Therefore we consider separately the cases with rotation added along $\phi$ and $\psi$. We refer to these two different singly spinning cases as Config.~A and Config.~B (see \fig{fig:ssdmp}). An outline of how to construct the solutions is given in \App{doubleMP}.

%figure

Consider first the solution described by Config.~A and fix the total mass so that we work with dimensionless quantities. When the total angular momentum, in this case $j_\phi$, is small enough, each black hole has its own ergosurface with $\Sp^3$ topology. When we vary  $j_\phi$ and the distance between the two black holes, the ergosurfaces can merge. The merger point is located at $\rho_0=0$, and, as expected, the tangents at the merger point have slopes $\sqrt{2}$ and the merger angle is $\theta_\rom{m} \sim 70.53^\circ$.  After the merger the topology of the ergosurface is $\Sp^2\times \Sp^1$. 
The ergosurface after merger remains pinned at the ``outer'' poles of the two black holes; this and related properties are discussed briefly in \App{topology}. The location of the merger point is on the ``$\psi$-axis'' where the orbits of $\partial_\psi$ shrink to zero. The merger ``surface'' is therefore 1-dimensional; so as for Saturn $\delta=3$.

In Config.~B the black holes have non-vanishing angular velocity $\Omega_\psi$, but $\Omega_\phi =0$. Only the first black hole has non-vanishing intrinsic (Komar) angular momentum $j_\psi$, but both black holes rotate and have ergoregions due to rotational frame dragging.
As $j_\psi$ is increased and the distance between the black holes decreased, the ergoregions can merge. The merger point is located at $\rho_0>0$, and hence the merger angle is $\theta_\rom{m} =\frac{\pi}{2}$. The topology of the ergosurfaces changes from two disjoint $\Sp^3$'s to an outer ergosurface of topology $\Sp^2\times \Sp^1$ inside which there is a second inner ergosurface, also of topology $\Sp^2\times \Sp^1$. The latter touches the black hole horizons on the poles where $\partial_\psi$ has fixed points. (See also \App{topology}.) Both the $\phi$- and $\psi$-circles have non-zero size at the merger point, so the merger surface is 2-torus of co-dimension $\delta=2$.

%____________________________________________
\section{Non-universality for multiply spinning configurations}
\label{nonunivex}
%____________________________________________

In this section we  consider ergosurface mergers in multiply spinning systems. We  focus our attention to five-dimensional Weyl-type solutions, where we can be fully explicit. For this class of doubly spinning solutions, we show that in general the angle at the merger point is not universal. To illustrate non-universality, we study ergosurface mergers in bi-rings and in the doubly spinning ring. This latter case is interesting by itself since it is the first example in which an ergosurface merges with itself. 

\subsection{Ergosurface mergers in five-dimensional doubly spinning Weyl solutions}

The reason why universality is lost for multiply spinning systems is that Einstein's equation for the $G_{tt}$ component of the metric no longer reduces the Laplace equation at the merger point. We illustrate this for the case of Weyl solutions with angular momentum in two planes.

The relevant component of Einstein's equation is given in \reef{Gttfineq}, and  
following our analysis in section \ref{weylergo}, we consider the expansion \reef{eqn:Gmerger} of $G_{tt}$ near the merger point. For simplicity, let us just consider the case where the merger point $(z_0,\rho_0)$ has $\rho_0>0$. Then the last term on the lhs of \reef{Gttfineq} vanishes, $G'_{tt}/\rho \to 0$, but the terms $(G'G^{-1}G')_{tt}$ and $(\ddot{G}G^{-1}\ddot{G})_{tt}$ have simple limits which can be expressed in terms of the off-diagonal components of the metric. We find
\begin{equation}
\label{merger2sa}
 a + b=\frac{1}{\rho_0^2}\Big[\big(G_{t\psi}\,G_{t\phi}'-G_{t\phi}\,G_{t\psi}'\big)^2
	+\big(G_{t\psi}\,\dot G_{t\phi}-G_{t\phi}\,\dot G_{t\psi}\big)^2\Big]\Big|_{(z,\rho)=(z_0,\rho_0)}\; .
\end{equation}
Clearly the rhs is positive (and non-zero) for a general doubly spinning configuration. (When the configuration is singly spinning we recover the result $a=-b$ found in \sec{weylergo} which implies that $\theta_\rom{m}= \pi/2$.)

In the doubly spinning case, the merger angle will depend on the values of $G_{t\phi}$ and $G_{t\psi}$ and their derivatives at the merger point, and we cannot expect universal behavior. The bi-ring system is an example of a doubly spinning solution for which the ergosurface merger points are located at $\rho_0>0$, and indeed we show in \sec{s:birings} that the merger angle depends on the parameters of the solution.

When $\rho_0=0$, the limit $(z,\rho) \to (z_0,\rho_0)$ is more subtle, as we already noted in \sec{weylergo}. An example of this type is the ergosurface self-merger of the doubly spinning black ring. We show in \sec{s:d2ring} that the merger angle for this system is non-universal.

%____________________________________________
\subsection{Bi-Rings as example of non-uniqueness}
\label{s:birings}
%____________________________________________

The 4+1-dimensional bi-ring solution describes two concentric black rings placed in orthogonal 2-planes. In the solution, constructed by the inverse scattering method in \cite{Izumi:2007qx,Elvang:2007hs}, each ring carries angular momentum in its respective plane. This allows balancing the solution, so that it is free of singularities on and outside the horizons. 

%figure
\begin{figure}[t]
\vspace{2cm}
\begin{center}
\begin{picture}(0,0)
\setlength{\unitlength}{1cm}
%lines and rods
\put(-5,1.5){\line(1,0){12}}
\put(-2.5,1.5){\linethickness{0.1cm}{\line(1,0){1.5}}}
\put(2.5,1.5){\linethickness{0.1cm}{\line(1,0){2}}}
\put(-5,0.5){\line(1,0){12}}
\put(-5,0.5){\linethickness{0.1cm}{\line(1,0){2.5}}}
\put(-1.0,0.5){\linethickness{0.1cm}{\line(1,0){1.75}}}
\put(-5,-0.5){\line(1,0){12}}
\put(0.75,-0.5){\linethickness{0.1cm}{\line(1,0){1.75}}}
\put(4.5,-0.5){\linethickness{0.1cm}{\line(1,0){2.5}}}
%labels
\put(-5.5,1.5){$t$}
\put(-5.5,-0.5){$\psi$}
\put(-5.5,0.5){$\phi$}
%rod directions
\put(-2.85,1.8){\small{$(1,\Omega^{(1)}_\phi,\Omega^{(1)}_\psi)$}}
\put(2.4,1.8){\small{$(1,\Omega^{(2)}_\phi,\Omega^{(2)}_\psi)$}}
\put(-4.5,0.75){\small{$(0,1,0)$}}
\put(-0.75,0.75){\small{$(0,1,0)$}}
\put(1.0,-0.25){\small{$(0,0,1)$}}
\put(5.0,-0.25){\small{$(0,0,1)$}}
%points
\put(-3.8,-1.1){$0$}
\put(-2.5,-1.1){$\ka_1$}
\put(-1.1,-1.1){$\ka_2$}
\put(0.6,-1.1){$\ka_3$}
\put(2.4,-1.1){$\ka_4$}
\put(4.35,-1.1){$\ka_5$}
\put(5.5,-1.1){$1$}
\end{picture}
\end{center}
\vspace{0.5cm}
\caption{\small{Rod structure of the bi-ring solution. The directions of the horizon rods depend on the angular velocities $\Omega^{(i)}_{\phi,\psi}$, which are given in (2.28)-(2.29) of \cite{Elvang:2007hs}.}}
\label{fig:biringrods}
\end{figure}
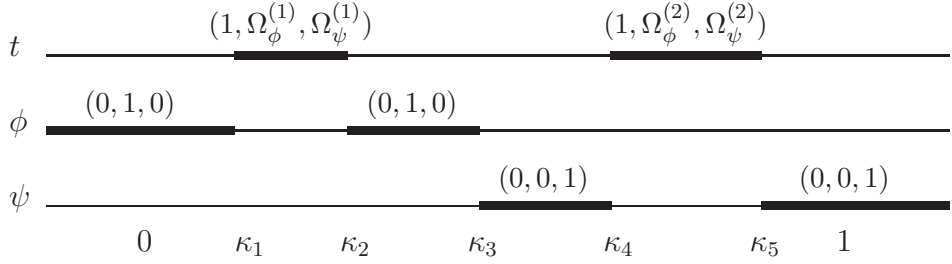

The properties needed for our analysis here can be found in \cite{Elvang:2007hs}, so we will be brief. The bi-ring solution is of the Weyl form and its rod configuration is shown in \fig{fig:biringrods}. In addition to an overall scale $L$, the solution is parametrized by five dimensionless parameters $\kappa_i$, $i=1,\dots 5$, which are directly related to the rod configuration in \fig{fig:biringrods} and satisfy the ordering 
\begin{equation}
 0< \kappa_1 < \kappa_2 < \kappa_3 < \kappa_4 < \kappa_5 < 1\;.
\end{equation}

The horizon of ring 1 is located at $\rho=0$ and $ z\in [\kappa_1,\kappa_2]$, and the horizon of ring 2 lies at $\rho=0$ and $z \in [\kappa_4,\kappa_5]$. The two ring planes are also located at $\rho=0$: for ring 1 the region $z\in ]\infty,\kappa_1]$ lies outside the ring and $z\in [\kappa_2,\kappa_3]$ inside, while for ring 2, $z\in [\kappa_3,\kappa_4]$ is inside the ring and $z\in [\kappa_5,\infty[$ outside.

The parameters $\kappa_i$ are constrained by two balance conditions (one for each ring):
\begin{eqnarray}
\begin{split}
&1~=~\frac{\sqrt{\kappa_3\kappa_5(1-\kappa_1)(\kappa_3-\kappa_2)(\kappa_4-\kappa_1)(\kappa_4-\kappa_2)(\kappa_5-\kappa_2)}}{\kappa_4(1-\kappa_2)(\kappa_3-\kappa_1)(\kappa_5-\kappa_1)}\;,\\[2mm]
&1~=~\frac{\sqrt{\kappa_5(1-\kappa_1)(1-\kappa_3)(\kappa_4-\kappa_1)(\kappa_4-\kappa_3)(\kappa_4-\kappa_2)(\kappa_5-\kappa_2)}}{\kappa_4(1-\kappa_2)(\kappa_5-\kappa_1)(\kappa_5-\kappa_3)}\;.
\end{split}
\label{bal12}
\end{eqnarray}
 
The conserved charges are the ADM mass $M$ and angular momenta $J_\phi$ and $J_\psi$ in the two planes of the rings. Fixing $M$, $J_\phi$ and $J_\psi$ leaves only one free parameter after solving the balance condition \req{bal12}, and hence the solution has 1-fold continuous non-uniqueness. This corresponds to different distribution of the mass between the two rings. Thus one can hold $M$, $J_\phi$ and $J_\psi$ fixed while varying the mass ratios of each ring $m_1$ and $m_2$ ($m_1+m_2=1$ are the Komar masses normalized by $M$). As the ring radii shrink and grow with changing mass distributions, their ergoregions enjoy a privilege to join and separate. 

\begin{figure}[t!]
 \begin{center}
 \begin{tabular}{ccl}
 \includegraphics[scale=0.85]{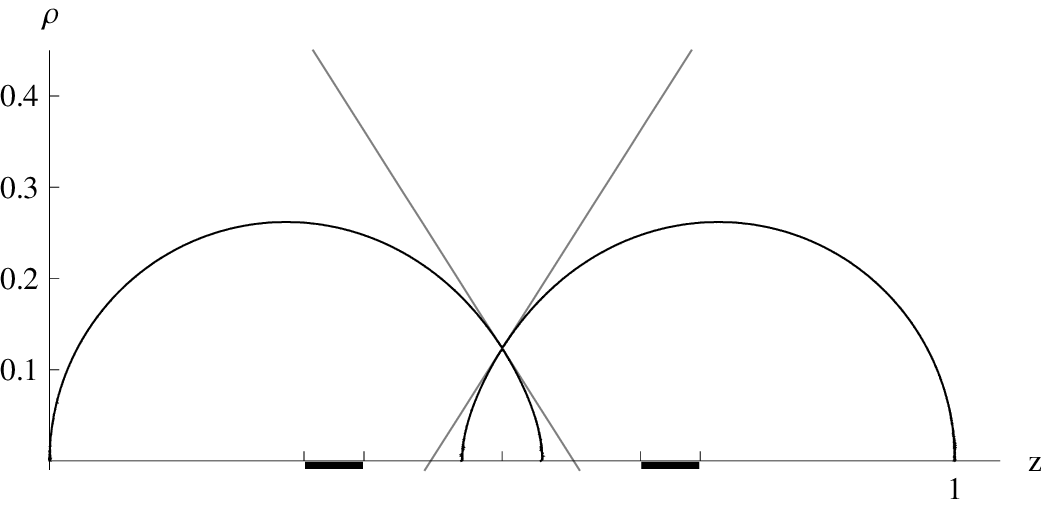} & ~ &
 \raisebox{2.1cm}{\small
   $\begin{array}{rcl}
   j_{1\psi}^2 &=& j_{2\phi}^2 ~=~0.24588\dots \\[1mm]
   m_1 &=&  m_2 ~=~ 1/2 \\[1mm]   
   \a_\pm &=& \pm 1.564\dots\\[1mm]  
   \theta_\textrm{m}&\sim&65.19^\circ
 \end{array}$}\\[2mm]
 {\footnotesize $\kappa_1 \approx 0.281\, , ~~
   \kappa_2 \approx 0.347\, , ~~
   \kappa_3 = 1/2\, , ~~
   \kappa_4 \approx 0.653\, , ~~
   \kappa_5 \approx 0.719$}  \\[4mm]
\includegraphics[scale=0.85]{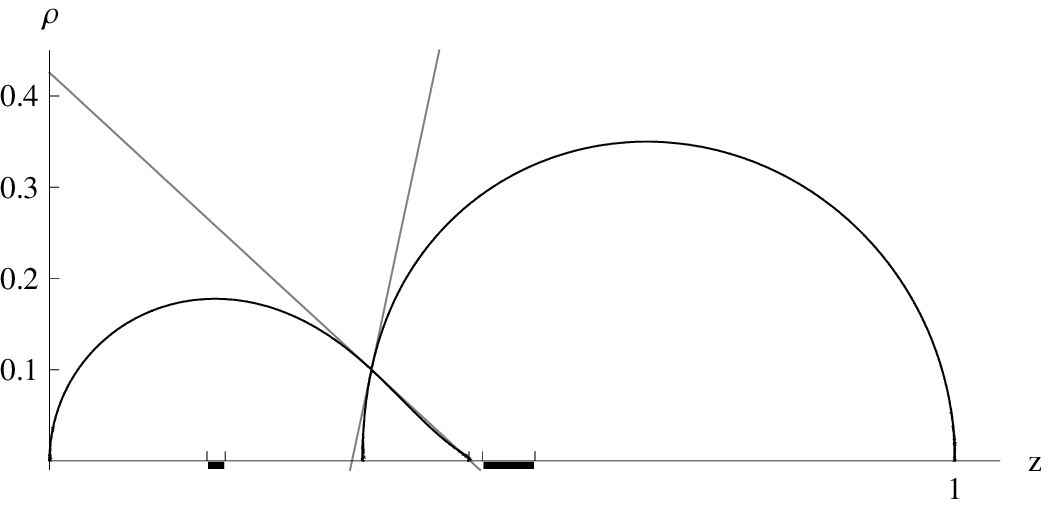} & ~~ &
 \raisebox{2.2cm}{\small
   $\begin{array}{rcl}
   j_{1\psi}^2 &=& 0.1\\[1mm] 
   j_{2\phi}^2 &=&0.45 \\[1mm]
   m_1 &=& 0.271\dots \\[1mm]
   \a_+ &=& 4.68\dots \\[1mm]
   \a_- &=& -0.916\dots  \\[1mm]
   \theta_\textrm{m}&\sim&59.57^\circ
 \end{array}$}\\[1mm]
    {\footnotesize 
    $\kappa_1 \approx 0.174\, , ~~
   \kappa_2  \approx 0.194\, , ~~
   \kappa_3  \approx 0.463\, , ~~
   \kappa_4  \approx 0.478\, , ~~
   \kappa_5  \approx 0.536$  } \\[4mm]
\includegraphics[scale=0.85]{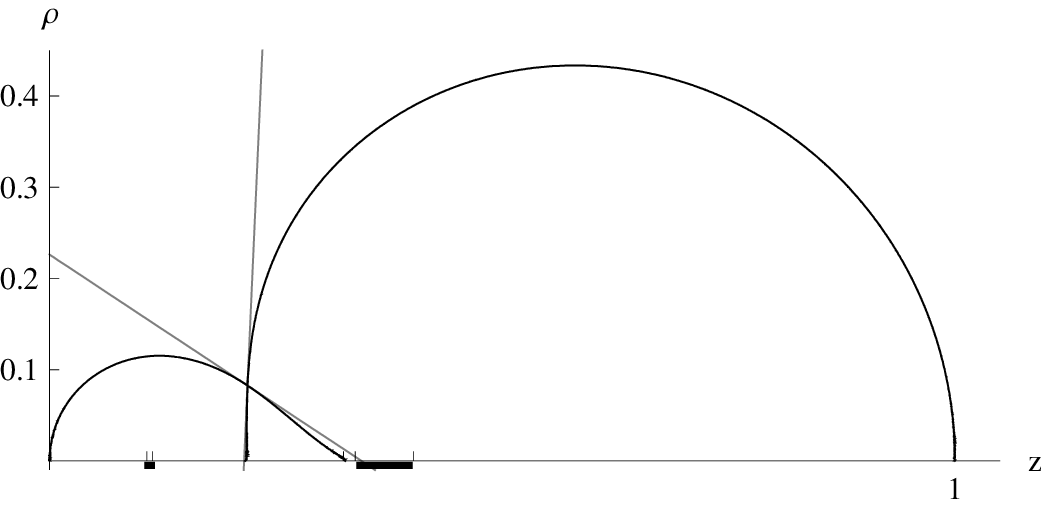} & ~~ &
 \raisebox{2.2cm}{\small
   $\begin{array}{rcl}
   j_{1\psi}^2 &=& 0.03\\[1mm]
   j_{2\phi}^2 &=& 0.65 \\[1mm]
   m_1 &=& 0.146\dots \\[1mm] 
   \a_+ &=& 22.29\dots \\[1mm]
   \a_- &=& -0.657\dots  \\[1mm]
   \theta_\textrm{m}&\sim& 59.26^\circ
 \end{array}$} \\[1mm]
    {\footnotesize 
    $\kappa_1 \approx 0.108\, , ~~
   \kappa_2 \approx 0.114\, , ~~
   \kappa_3 \approx 0.325\, , ~~
   \kappa_4 \approx 0.337\, , ~~
   \kappa_5 \approx 0.402$}   
\end{tabular}
\end{center}
\caption{Non-uniqueness in bi-ring ergoregion mergers. The ergosurface locii are shown at the merger points for three different sets of values of parameters. We fix the conserved angular momenta and the total mass and tune the balanced bi-ring solution to be at the ergosurface merger point. Tangents at the merger point are included (in gray) to guide the eye. The value of the merger angle $\theta_\rom{m}$ is given in each case.
The ticks on the $z$-axis mark the values of the $\kappa_i$. The horizontal bars indicate the location of the horizons.}
\label{fig:mergersBiRing}
\end{figure}

We show three examples of bi-ring ergoregion mergers in \fig{fig:mergersBiRing}. The first case shows the ergosurface merger for the symmetric bi-ring configuration, which was studied in detail in \cite{Elvang:2007hs}. In this case the rings are identical, so the magnitudes of the angular momenta in the two planes are the same, and they satisfy $j^2 > 0.246$. The phase diagram is shown in Fig.~5(a) of \cite{Elvang:2007hs}. The ergoregion merger happens at $j^2 \approx 0.24588$ and takes place on the fat ring branch. The slopes of the tangents at the merger point are $\alpha_\pm \approx \mp 1.564$.

The two other examples of ergoregion mergers in \fig{fig:mergersBiRing} are for asymmetric bi-ring configurations. We fix the dimensionless angular momenta at different values, then solve the balance conditions and merger conditions all simultaneously to find the values of $\ka_i$ and $(z_0,\rho_0)$ at the merger point. The freedom to find the merger point lies in the 1-fold non-uniqueness allowing different mass distributions. In each case in \fig{fig:mergersBiRing} we give the corresponding mass ratio $m_1$ for ring 1. 

It is visually obvious that the tangents at the merger points have different slopes in the three examples of \fig{fig:mergersBiRing}. However, it is more important to realize that the angle between the tangents changes as the parameters are varied. Thus the merger angle depends in detail on the parameters of the solution and there is no universality.  

Finally, let us point out that when the ergosurfaces merge in bi-ring configurations, the topology changes from two disjoint $\mathbf{S}^2\times\mathbf{S}^1$ to $\mathbf{S}^3$. To be more precise, after the merger, there are both an inner and an outer ergosurface of topology $\mathbf{S}^3$: the black ring horizons both lie inside the outer one, but outside the inner one. The surface of merger points is a 2-torus.

%____________________________________________
\subsection{Doubly spinning black ring as an example of self-merger}
\label{s:d2ring}
%____________________________________________

The doubly spinning ring \cite{Pomeransky:2006bd} has angular momentum in the plane of the ring, as needed for balance, and in the orthogonal plane, i.e.~the $\mathbf{S}^2$ of the ring is also rotating. When the ring is large and thin, the ergosurface has topology $\mathbf{S}^2 \times \mathbf{S}^1$, just like a singly spinning ring, but when the ring becomes fatter, the spin of the $\mathbf{S}^2$ plays a significant role. It turns out that the ergoregion can merge with itself across the center of the ring, so that the topology of the ergosurface changes from $\mathbf{S}^2 \times \mathbf{S}^1$ to $\mathbf{S}^3$.\footnote{This has also been noticed by Mark Durkee (private communication).}  At the same time, an inner $\mathbf{S}^3$ ergosurface appears to exclude the center of the ring from the ergoregion. This is necessary because the center in the plane of the ring cannot belong to the ergoregion as it is a point of symmetry.
For comparison, the singly spinning black ring with $\mathbf{S}^1$ angular momentum only, has an ergosurface with topology $\mathbf{S}^2 \times \mathbf{S}^1$ which never self-intersects, except in the limit where the fat ring becomes singular as $j\to 1$ and $a_\rom{H} \to 0$.\footnote{We are grateful to R.~Emparan for discussions of this and several other related points.}

Let us first briefly review the balanced doubly spinning black ring solution of  \cite{Pomeransky:2006bd}. The metric is written in ring coordinates $(x,y)$ with $-1\le x \le 1$ and $y\le -1$ and $(x,y) \to (-1,-1)$ being asymptotic infinity (see \cite{Emparan:2001wn} for further details). The part of the plane of the ring connecting to the outer rim of the ring is located at $x=-1, \, y \le -1$, while the part of the plane connecting to the inner rim is at $x=+1, \, y \le -1$. 

The solution has one dimensionfull parameter $k$, which sets the scale, and two dimensionless parameters $\lambda$ and $\nu$ satisfying
\bea
  \label{lamnu}
  0\leq \nu <1\,  \hspace{1cm}
  2\sqrt\nu\leq\lambda<1+\nu \, .
\eea
The limit $\lambda \to 2\sqrt{\nu}$ gives an extremal ring with zero temperature; a microscopic calculation of its entropy was presented in \cite{Reall:2007jv}. In the limit $\lambda \to 1+\nu$ the ring collapses to an extremal Myers-Perry black hole \cite{Reall:2007jv,Elvang:2007hs}.

The $tt$ component of the doubly spinning black ring metric \cite{Pomeransky:2006bd} is
\begin{equation}
 G_{tt}=-\frac{H(y,x)}{H(x,y)}\,, \label{eqn:gttD2ring}
\end{equation}
where 
\bea
\label{Hyx}
H(y,x)=1+\lambda ^2-\nu ^2+2\lambda\,\nu\, \left(1-y^2\right) x+2\lambda  \left(1-x^2 \nu ^2\right)y  + \nu  \left(1-\lambda ^2-\nu ^2\right)x^2 y^2\, .
\eea
The ergosurface is located at the solutions to $H(y,x)=0$.

%figure
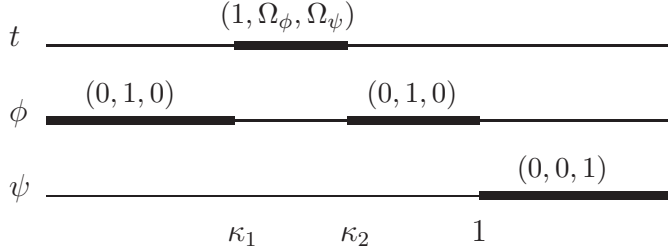
\begin{figure}[t]
\vspace{2cm}
\begin{center}
\begin{picture}(0,0)
\setlength{\unitlength}{1cm}
%lines and rods
\put(-5,1.5){\line(1,0){8.25}}
\put(-2.5,1.5){\linethickness{0.1cm}{\line(1,0){1.5}}}
\put(-5,0.5){\line(1,0){8.25}}
\put(-5,0.5){\linethickness{0.1cm}{\line(1,0){2.5}}}
\put(-1.0,0.5){\linethickness{0.1cm}{\line(1,0){1.75}}}
\put(-5,-0.5){\line(1,0){8.25}}
\put(0.75,-0.5){\linethickness{0.1cm}{\line(1,0){2.5}}}
%labels
\put(-5.5,1.5){$t$}
\put(-5.5,-0.5){$\psi$}
\put(-5.5,0.5){$\phi$}
%rod directions
\put(-2.7,1.8){\small{$(1,\Omega_\phi,\Omega_\psi)$}}
\put(-4.5,0.75){\small{$(0,1,0)$}}
\put(-0.75,0.75){\small{$(0,1,0)$}}
\put(1.25,-0.25){\small{$(0,0,1)$}}
%points
\put(-2.6,-1.1){$\ka_1$}
\put(-1.1,-1.1){$\ka_2$}
\put(0.65,-1.1){$1$}
\end{picture}
\end{center}
\vspace{0.5cm}
\caption{\small{Rod structure of the doubly spinning ring. Expressions for the angular velocities $\Omega_{\phi,\psi}$ in terms of $\lambda$ and $\nu$ can be found in \cite{Elvang:2007hs}. The relationship between $\kappa_{1,2}$ and $\lambda, \nu$ is given in \req{D2kappa}.}}
\label{fig:D2ringrods}
\end{figure}

Let $\partial_\psi$ be the Killing vector generating the $\mathbf{S}^1$ of the ring. The orbits of $\partial_\psi$ close off smoothly at $y=-1$, which we will denote the ``$\psi$-axis". (We use the notation of \cite{Elvang:2007hs} where $\psi$ parameterizes the $\mathbf{S}^1$ of the ring and $(x,\phi)$ parameterize the $\mathbf{S}^2$.) If the ergosurface is to intersect itself, the merger must necessarily take place on the $\psi$-axis. Setting $y=-1$ in \req{Hyx} we find that the ergosurface intersects the $\psi$-axis at
\bea
  x_\pm = \pm \sqrt{\frac{\lambda+\nu-1}{\nu\,(1+\lambda-\nu)}} \, . 
\eea
The expression under the squareroot is non-negative when $\lambda+\nu\ge 1$, so only such solutions can have self-merging ergoregions. It can be verified that $0\le  x_\pm^2 \le 1$ for all $\nu,\lambda$ satisfying $\lambda+\nu\ge 1$ in addition to \req{lamnu}. The solution $x_-$ ($x_+$) is the intersection of the outer (inner) ergosurface with the $\psi$-axis. The merger point is where the inner and outer ergosurfaces precisely touch, and that happens when $x_+ = x_- = 0$, i.e.~when 
\bea
  \lambda_\rom{m} = 1-\nu \, .
\eea
The lower bound \req{lamnu} on $\lambda$ implies that a merger point only exists when $\nu \le 3 - 2\sqrt{2} \approx 0.1716$.

In the previous sections we characterized the merger point as a point where $G_{tt}$ and its first derivatives vanished. The merger point $(x,y)=(0,-1)$ identified above satisfies precisely these conditions, but we have to be careful when verifying it. This is because the metric has a coordinate singularity at $y=-1$, where the metric component $G_{yy} \propto \frac{1}{1-y^2}$ diverges. Introducing a new coordinate $Y$ by setting $y=-1-Y^2$, the metric will be regular at $Y=0$. It is simple to verify then that $G_{tt} = \partial_x G_{tt} = \partial_Y G_{tt} = 0$ has only one solution, namely $(x,Y)=(0,0)$ and $\lambda = \lambda_\rom{m} = 1- \nu$.

%figure
\begin{figure}[t]
%\begin{center}
\includegraphics[scale=0.9]{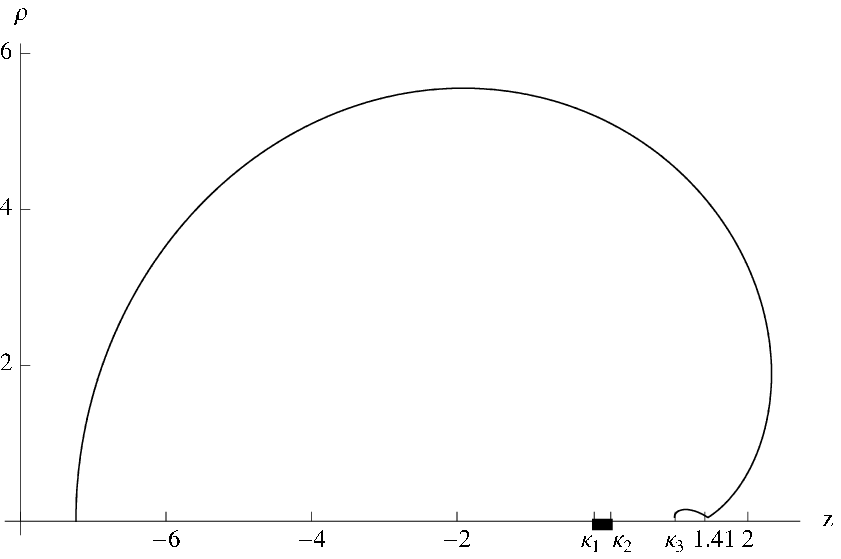}
\hspace{-6.5cm}
\raisebox{0.8cm}{\hspace{0.3cm}
  \includegraphics[width=3.cm]{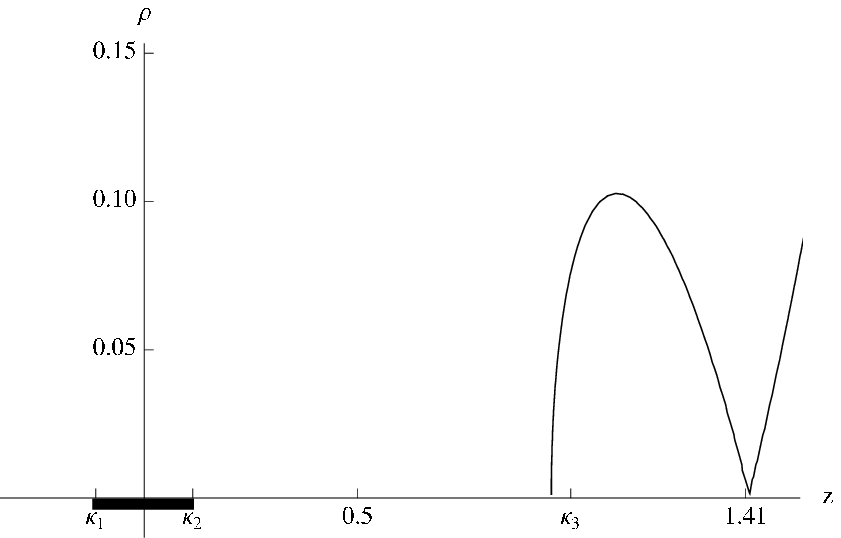}}
\hspace{4cm}
\includegraphics[scale=0.9]{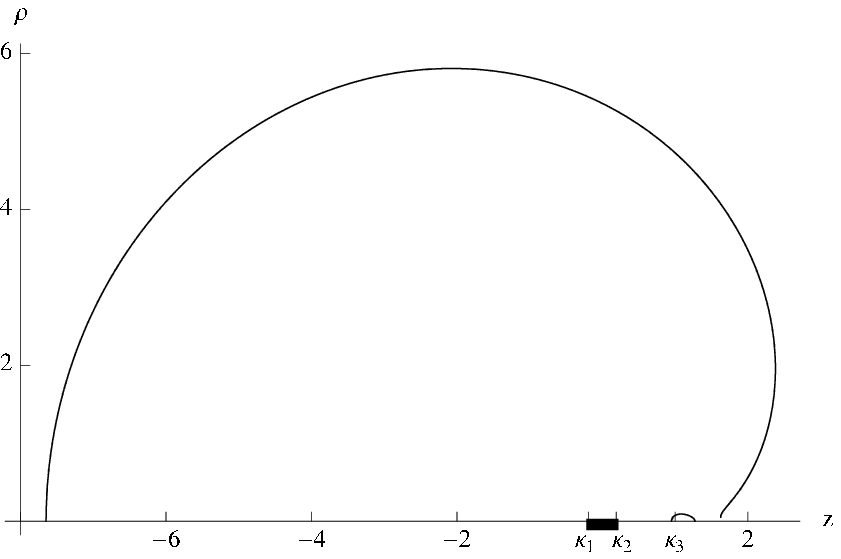}
\hspace{-6.5cm}
\raisebox{0.8cm}{\hspace{0.2cm}
  \includegraphics[width=3.cm]{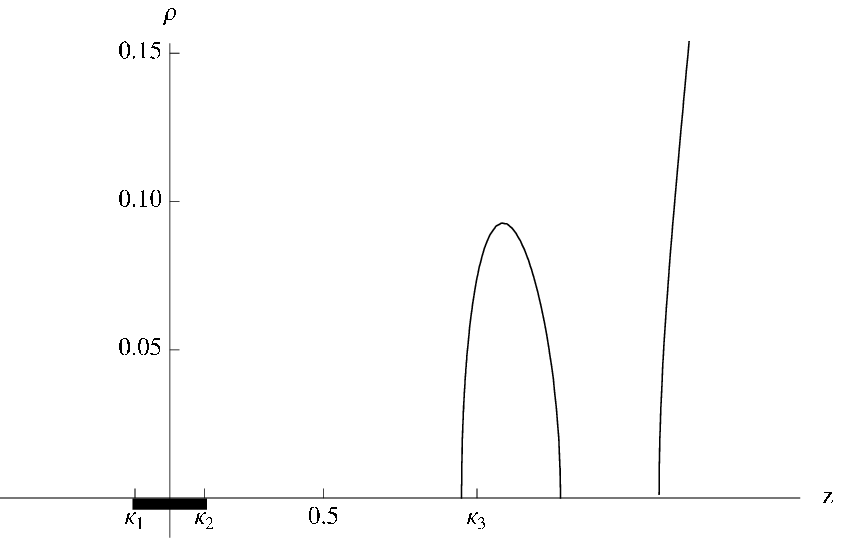}}
%\end{center}
\caption{\textit{Left:} Self-merger of the ergosurface in the doubly spinning ring with $\nu=0.17$. We have fixed the scale by setting $k=1$. The superimposed figure shows in detail the region of interest. The merger angle is $\theta_\rom{m} \approx 119.5^\circ$. \textit{Right:} As the ring becomes fatter, $\lambda>\lambda_c$, there are two ergosurfaces of $\mathbf{S}^3$ topology, and inner and an outer one.}
\label{fig:dspinring}
\end{figure}

To be able to compare directly with the other solutions considered in this paper, we will now switch to Weyl coordinates $(z,\rho)$. The needed coordinate transformation from ring coordinates $(x,y)$ is given in \App{weyldspinring}. Figure \ref{fig:D2ringrods} shows the rod structure; the rod endpoints $\kappa_{1,2}$ are related to $\lambda, \nu$ by
\begin{equation}
\kappa_1=-\frac{\sqrt{\lambda-4\,\nu^2}}{1-\nu}\,,\qquad \kappa_2=\frac{\sqrt{\lambda-4\,\nu^2}}{1-\nu} \;.
\label{D2kappa}
\end{equation}

By the coordinate transformation given in \req{eqn:xyD2ring}, the merger point is located at $\rho_0=0$ and $z_0=\frac{1+\nu}{1-\nu}$. (We have fixed the scale by setting $k=1$.) Expanding $G_{tt}$ near the merger point gives
\begin{equation}
G_{tt}\approx -\frac{1-\nu}{8}(z-z_M)^2+\frac{1-\nu}{16\,\nu}\,\rho^2+\dots
\end{equation}
Hence, the slopes of the tangents at the merger point are $\pm \sqrt{2\nu}$, and as a consequence the merger angle $\theta_\rom{m}=2\,\rom{arccot}\,(\sqrt{2\nu})$ is non-universal.

Figure \ref{fig:dspinring} shows an example of the ergosurface locii merger in the $(z,\rho)$ plane. This is the case where $\nu=0.17$, so the merger angle is $\theta_\rom{m} \approx 119.5^\circ$. Generally we find that the merger angle varies between $119.28^\circ \lesssim \theta_\rom{m} <180^\circ$. The lower bound is obtained when $\nu \to 3-2\sqrt{2}$, whereas $\theta_\rom{m}$ approaches $180^\circ$ in the limit where the $\mathbf{S}^2$ angular momentum vanishes and the ergosurface only merges in the singular limit.

\begin{figure}[t]
\begin{center}
 \includegraphics[scale=1.3]{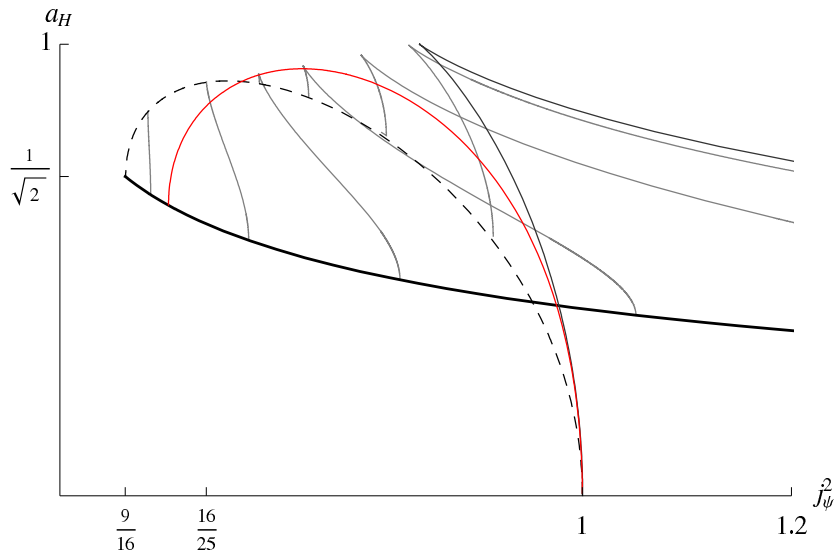}
\end{center}
\caption{Self-merger phases in the doubly spinning ring phase diagram (see \cite{Elvang:2007hs} for further details). 
The \textit{light gray} curves correspond to the phases of doubly spinning black rings with constant $j_\phi$, for $j_\phi^2=\frac{1}{18},\frac{1}{25},\frac{1}{35},\frac{1}{50},\frac{1}{100},\frac{1}{500}$ (from left to right).
The \textit{thin black} curve shows the phase of singly spinning black ring, and the \textit{thicker black} curve shows the phase of extremal doubly spinning ring. The red curve shows where the self-merger of the ergoregions takes place. There is at most one self-merger configuration for each $j_\phi$. The \textit{dashed} curve is the limit where the ring collapses to an extremal Myers-Perry black hole; the horizon area is discontinuous in this limit, so the dashed curve does not itself represent a phase, but only outlines the limiting value of the black ring area.}
\label{fig:D2phases}
\end{figure}

Where in the phase diagram $a_\rom{H}$ vs.~$j_\psi^2$ does the ergoregion mergers take place? We illustrate this in \fig{fig:D2phases}. The gray curves are doubly spinning black rings with fixed $j_\phi$, which can vary between $0$ and $1/4$. These curves begin on the monotonically decreasing black curve, which is the phase of the extremal doubly spinning black ring, and they end on the dashed curve; in this limit the ring collapses to an extremal Myers-Perry black hole.\footnote{The horizon area jumps discontinuously in this limit, so the dashed curve only represents the limiting value of the doubly spinning ring's area, it does not represent the actually phase of the extremal Myers-Perry black hole.} The red curve in \fig{fig:D2phases} shows where on the fixed-$j_\phi$ branches the mergers occur. When $j_\phi$ is sufficiently large, 
$j_\phi > j_{\phi,c} \equiv \frac{1}{2}\sqrt{\frac{1}{\sqrt{2}} -\frac{1}{2}}\approx 0.228$, the ergoregion does not undergo self-merger, but it is so large that it always connects across the $\psi$-axis. In that case there is an inner and outer ergosurface, each of topology $\mathbf{S}^3$. Note that $j_{\phi,c}$ is the value of $j_\phi$ for which the ergosurface of the extremal doubly spinning black ring self-intersects. The merger surface is 1-dimensional, so $\delta = 3$.

%____________________________________________
\section{Discussion}
\label{discuss}
%____________________________________________

We have analyzed key aspects of ergosurface mergers, uncovering an intriguing
universality for a certain wide class of stationary spacetimes.  We considered
asymptotically flat vacuum solutions with rotation in a single plane that depend on two variables:  these can be thought of as a radius in the plane and a distance off the plane. (We also assumed a discrete reflection symmetry.) This includes all known singly spinning black hole solutions. We proved that if ergosurfaces merge, the merger angle is universal: it depends only on whether the merger occurs on the plane of rotation or off, not on  any specific details such as mass and angular momentum. This remarkable result follows from the structure of Einstein's equation.  Indeed, the relevant component of Einstein's equation becomes the Laplace equation for $G_{tt}$ near the merger point, so the merger behaviour is mimicked
by many more familiar systems in nature, such as Newtonian equipotentials.

To demonstrate that our results pertain to known solutions, namely systems of black holes and black rings, we have analyzed a number of such solutions explicitly. Table 1 summarizes the examples of ergosurface mergers discussed in Sections \ref{univex} and \ref{nonunivex}. All examples of exact black hole solutions where we have investigated ergoregion mergers belong to the Weyl class. The location of the merger point, whether at $\rho =0$ or $\rho > 0$, is in examples of regular Weyl solutions related to the co-dimension $\delta$ of the merger surface, and the merger angle is then simply $\theta_\rom{m} = 2\,\rom{arccot}\,\sqrt{\delta-1}$. (For nakedly singular Weyl solutions, or more generally for the solutions considered in \sec{hdmergers}, we do not have a proof of the relationship between the location of the merger point at $\rho =0$ or $\rho > 0$ and the co-dimension of the merger surface.)

\begin{table}[thdp]
\begin{center}
\begin{tabular}{|l|c|c|c|c|c|c|c|}\hline
 Weyl solution & ~$J$~ &  balanced & merger point & ~$\delta$~ &$\theta_\rom{m}$  \\ \hline 
4d double Kerr & 1 & no & $\rho_0 > 0$ &2& $90^\circ$ \\
Black Saturn / di-ring & 1 & yes & $\rho_0 = 0$ &3& $2\,\rom{arccot}\,\sqrt{2} \to 70.53^\circ$ \\
5d double MP (A) & 1 & no & $\rho_0 = 0$&3& $2\,\rom{arccot}\,\sqrt{2}  \to 70.53^\circ$ \\
5d double MP (B) & 1 & no & $\rho_0 > 0$ &2& $90^\circ$ \\
singly-spinning Bi-ring & 1 & no & $\rho_0 > 0$ &2& $90^\circ$ \\
Bi-ring  & 2 & yes & $\rho_0 > 0$ &2& non-universal \\ 
doubly-spinning BR & 2 & yes & $\rho_0 = 0$ &3& $119.28^\circ \lesssim \theta_\rom{m} <180^\circ$ \\
 \hline 
 \end{tabular} 
 \caption{Summary of known results for the merger slope in the $(z,\rho)$ plane for various configurations. The column $J$ shows in how many independent planes the solution has rotation. $\delta$ is the co-dimensionality of the merger surface.}
\end{center}
\label{summarytab}
\end{table}

Table 1 demonstrates explicitly that single spin is crucial  to guarantee universality. It is clear that  we need not require that the solution in question is balanced; conical singularities do not interfere with universality. On the other hand, ergosurfaces of doubly-spinning solutions merge in a non-universal fashion.  In the case of multiply spinning systems, as we have seen explicitly in  \sec{nonunivex}, the lack of universality arises due to interactions between spins in different planes. The spins conspire to induce a source term for the equation of motion for $G_{tt}$ near the merger point. As the source depends on the physical details of the constituent black holes, universality is lost.

Studying ergosurface mergers rewarded us with a rich set of examples of topology change for physically preferred surfaces. Unlike the situation for black hole horizons, whose allowed topology is restricted by the topology theorems, there are no known constraints on ergosurface topology.  For example, already in 4 spacetime dimensions a toroidal ergosurface is possible, albeit for nakedly singular configurations.   More intriguingly, as exemplified by the doubly spinning black ring,   one can obtain a nested set of ergosurfaces, i.e.~a system with an inner and outer ergosurface, both of which are located outside the black hole horizon. 

A question regarding multi-ring solutions is whether it is possible to engineer solutions where multiple ergosurfaces merge simultaneously. Consider for example Saturn with two rings: when the constituents are far apart, there are three disjoint ergosurfaces with topologies $\Sp^3$ (for the Myers-Perry black hole) and $\Sp^2 \times \Sp^1$ (for the rings). With both rings in the same plane and angular momentum only in that plane, presumably the solution parameters can be tuned so that the ergosurfaces merge pairwise at the same values of parameters. Perhaps more interesting is a setup with the two rings in orthogonal planes. With angular momentum now in both planes of rotation, it may be possible to choose parameters such that all three ergosurfaces merge at the same point. 
It would also be interesting to understand better the role of inner ergoregions (located inside another ergoregion but outside any event horizons). In our examples, inner ergosurfaces seem to be present when needed to prevent a point of symmetry, like the center of a black ring, from being part of an ergoregion.

The physical importance of ergoregions motivated us to consider properties of merging ergosurfaces, defined with respect to asymptotic static observers.  However, we could have considered analogous surfaces defined with respect to a set of distant orbiting observers\footnote{We thank Bill Unruh for discussions on this point.}, or phrased mathematically, the surface given by vanishing norm of a Killing field
$\xi^a \equiv (\partial_t)^a + \a (\partial_\psi)^a$ for some constant $\a$.
The local nature of our universality proof would then imply that the surfaces of $\xi^a \, \xi_a = 0$ would likewise merge at the same universal angle.  This implies that universality does not just pertain to a very finely-tuned\footnote{The fine-tuning refers to the fact that ergoregion mergers occur at a specific locus in parameter space.} set of solutions, but rather to  a full open set (in the parameter space) of configurations. This implies a specific rigid structure of a large set of stationary spacetimes in general relativity.

It is worth noting that in the static limit of no rotation, the equation defining the location of the ergoregion becomes the condition for the presence of a horizon. Might our results for merging ergoregions then teach us something about black hole mergers, such as the black hole - black string transition? Unfortunately, this is not the case, since the merger point in such a transition is a curvature singularity. 

In a spacetime with an ergoregion we expect to see superradiant
scattering, i.e.~a wave impinging on a rotating black hole will re-emerge with a larger amplitude. This is essentially a consequence of the asymptotic
timelike Killing field turning spacelike in the ergoregion. The modes involved in superradiant scattering have low frequency and long wavelength. It is interesting to ask whether the mergers of ergosurfaces leave an imprint on the superradiance pattern. A priori the answer  would seem to be yes, as can be seen by the following gedanken experiment: suppose we send such a wave into our system. If there are two disconnected ergoregions, one might expect the reflected wave to show interference patterns. After the ergoregions
merge, one does not expect interference. It would be interesting to analyze this further.

%%%%%%%%%%%%%%%%%%%%%%%%%%%%%%%%%%%%%%%%%%%%%%%%%%
%____________________________________________
\subsection*{Acknowledgements}
\label{acks}

It is a pleasure to thank Piotr Chru\'sciel, James Lucietti, Harvey Reall, Simon Ross and especially Roberto Emparan for useful discussions. 

VH and MR would like to thank MIT for hospitality during the initial stages of this project. In addition, VH and MR would also like to thank the Newton Institute, the Galileo Galilei Institute, the Tata Institute for Fundamental Research,  University of British Columbia, CERN, and  IPMU for hospitality during the course of this project.  

HE is supported by NSF grant PHY-0503584. During this work, HE was supported by a Pappalardo Fellowship in Physics at MIT, and in part by the US Department of Energy through cooperative research agreement DE-FG0205ER41360. PF is supported by STFC. GH is supported in part by NSF grant PHY-0555669.   VH and MR were supported in part by STFC and by INFN. 

%____________________________________________
\appendix

%____________________________________________
\section{Einstein's equation with two commuting Killing fields}
\label{geroch2}
%____________________________________________

In this appendix we study Einstein's equation in $d+1$ dimensions assuming the existence of two commuting Killing vector fields by generalizing the construction of \cite{Geroch:1970nt,Geroch:1972yt}.

Since the two Killing fields commute, it is possible to choose adapted coordinates so that $\xi^a=(\partial_t)^a$ is the timelike Killing vector at infinity and $\psi^a=(\partial_\psi)^a$ is the generator of rotations. We shall further assume that $(t,\psi)\leftrightarrow (-t,-\psi)$ is a symmetry of the spacetime, which is a reasonable physical requirement for any rotating body. 

Consider the inner products of the Killing fields,
\begin{equation}
\label{Bscalars}
\G_{tt}=\xi^a\xi_a\,,\qquad 
\G_{t\psi}=\xi^a\psi_a\,,\qquad
\G_{\psi\psi}=\psi^a\psi_a\,.
\end{equation}
Notice that for a Lorentzian spacetime we have $\G_{tt}<0$ near infinity. Then, the general metric satisfying our assumptions is given by
\begin{equation}
 ds^2= \G_{tt}\,dt^2+2\,\G_{t\psi}\,dt\,d\psi+\G_{\psi\psi}\,d\psi^2+ds^2(\CB)\,,
\end{equation}
where $\CB$ is a Riemannian manifold, which we refer to as the base. It is useful to define a matrix of the scalar fields \reef{Bscalars} on the base $\CB$ 
\begin{equation}
\label{gamma}
 \G=\left(
\begin{array}{cc}
\G_{tt} & \G_{t\psi}\\
\G_{t\psi} & \G_{\psi\psi} 
\end{array}
\right)\,,\qquad \tau =-\det\G\,. 
\end{equation}
The metric becomes degenerate if $\tau = 0$. We are restricting the analysis to non-singular regions of the spacetime where $\tau$ is non-vanishing.

The metric on the base $\CB$ can be expressed in terms of the spacetime metric $g_{ab}$ and the data involving the Killing fields $\xi^a$ and $\psi^a$ as 
\begin{equation}
 h_{ab}=g_{ab}+\frac{1}{\tau}\left[\G_{\psi\psi}\,\xi_a\,\xi_b+\G_{tt}\,\psi_a\,\psi_b	-2\,\G_{t\psi}\,\xi_{(a}\psi_{b)})\right]\,.
\end{equation}

The derivatives of the Killing fields can be computed directly as follows:
\begin{subequations}
 \begin{align}
  \nabla_a\xi_b&=\frac{1}{\tau}\left[\xi_{[a}\big(\G_{\psi\psi}\,D_{b]}\G_{tt}-\G_{t\psi}\, D_{b]}\G_{t\psi}\big)
	-\psi_{[a}\big(\G_{t\psi}D_{b]}\G_{tt}-\G_{tt}D_{b]}\G_{t\psi}\big)\right]\,,\\
 \nabla_a\psi_b&=\frac{1}{\tau}\left[\psi_{[a}\big(\G_{tt}\, D_{b]}\G_{\psi\psi}-\G_{t\psi}\, D_{b]}\G_{t\psi}\big)
	-\xi_{[a}\big(\G_{t\psi}\, D_{b]}\G_{\psi\psi}-\G_{\psi\psi}\, D_{b]}\G_{t\psi}\big)\right]\,.
 \end{align}
\label{eqn:dxi}
\end{subequations}
In these expressions $D_a$ is the covariant derivative on $\CB$ associated to the metric $h_{ab}$, and $\nabla_a$ is the covariant derivative associated to the full spacetime metric $g_{ab}$. 

We now derive the first set of the Einstein's equations for the components of the matrix $\G$. We compute directly
\begin{eqnarray}
 D^aD_a\G_{tt}&=&h^{ab}\nabla_a\left(h_b^{\phantom b m}\nabla_m\G_{tt}\right) \nonumber\\
	&=&2\,h^{ab}\nabla_a\left(\xi^m\nabla_b\xi_m\right) \nonumber \\
	&=&2\,h^{ab}\xi^m\nabla_a\nabla_b\xi_m+2\,h^{ab}(\nabla_a\xi^m)(\nabla_b\xi_m) \nonumber \\
&=&-2\,R_{mn}\,\xi^m\xi^n-\frac{2\,\G_{tt}}{\tau}\,R_{mnpq}\,\xi^m\psi^n\xi^p\psi^q \nonumber \\
&&\hspace{0.5cm}	-\frac{1}{2\,\tau}\left[\G_{\psi\psi}(D\G_{tt})^2+\G_{tt}(D\G_{t\psi})^2
	-2\,\G_{t\psi}(D^a\G_{tt})(D_a\G_{t\psi})\right]\,.
\label{eqn:pD2l00}
\end{eqnarray}
In the second step we use that the Killing vectors commute. The last equality follows from \eqref{eqn:dxi} and the fact that a Killing field satisfies
\begin{equation}
 \nabla_a\nabla_b\xi_c=R_{dabc}\,\xi^d\,.
\label{eqn:Riemann}
\end{equation}
Using \eqref{eqn:Riemann}, one can show that
\begin{equation}
 R_{mnpq}\,\xi^m\psi^n\xi^p\psi^q=-\frac{1}{4}\left[(D^a\G_{tt})(D_a\G_{\psi\psi})-(D\G_{t\psi})^2\right]\,,
\end{equation}
and hence \eqref{eqn:pD2l00} becomes
\begin{equation}
\begin{aligned}
  D^aD_a\G_{tt}&=-2\,R_{mn}\,\xi^m\xi^n
	+\frac{\G_{tt}}{2\,\tau}\left[(D^a\G_{tt})(D_a\G_{\psi\psi})-2\,(D\G_{t\psi})^2\right]\\
&\hspace{0.5cm}-\frac{1}{2\,\tau}\left[\G_{\psi\psi}(D\G_{tt})^2-2\,\G_{t\psi}(D^a\G_{t\psi})(D_a\G_{tt})\right]\,.
\end{aligned}
\label{eqn:D2l00}
\end{equation}

The equations for the remaining metric components $\G_{t\psi}$ and $\G_{\psi\psi}$ can be derived analogously. This set of  equations can be cast into a compact form using the matrix structure. One has 
\begin{equation}
\begin{aligned} 
 D^aD_a\G_{\alpha\beta}&=-2\,R_{mn}\,\xi_{(\alpha)}^m\xi_{(\beta)}^n
	+(\G^{-1})^{\mu\nu}(D^a\G_{\alpha\mu})(D_a\G_{\beta\nu})
	-\frac{1}{2\,\tau}\,(D^a\tau)(D_a\G_{\alpha\beta})\,,
\end{aligned}
\label{eqn:scalar}
\end{equation}
where we have defined $\xi^a_{(t)}=\xi^a$ and $\xi^a_{(\psi)}=\psi^a$ in order to simplify the notation, and  matrix multiplication is understood.

To derive the second set of the Einstein equations, we proceed as in \cite{Geroch:1972yt} and consider an arbitrary vector $k^c$ on $\CB$. By evaluating the commutator of covariant derivatives on this one-form we can extract the Ricci tensor on the base $\CB$; starting from 
\begin{equation}
 \begin{aligned}
  D_{[a} D_{b]}k_c&=h_{[a}^{\phantom{[a}m}h_{b]}^{\phantom{b]}n}h_{c}^{\phantom c p}\,
	\nabla_m(h_{n}^{\phantom n r}h_{p}^{\phantom p s}\nabla_r k_s)\\
&=h_{a}^{\phantom a p}h_{b}^{\phantom b q}h_{c}^{\phantom c r}\,\nabla_{[p}\nabla_{q]} k_r\\
&\hspace{0.5cm}-\frac{1}{\tau}\,h_{a}^{\phantom a m}h_{b}^{\phantom b n}h_{c}^{\phantom c p}\big[
	\G_{\psi\psi}(\nabla_p\xi^r)(\nabla_m \xi_n)+\G_{tt}(\nabla_p\psi^r)(\nabla_m\psi_n)\\
&\hspace{3.5cm}-\G_{t\psi}\big(
	(\nabla_p\xi^r)(\nabla_m \psi_n)+(\nabla_p\psi^r)(\nabla_m \xi_n)
\big)\big]k_r\\
&\hspace{0.5cm}-\frac{1}{\tau}\,h_{[a}^{\phantom{[a} m}h_{b]}^{\phantom{b]} n}h_{c}^{\phantom c p}\big[
	\G_{\psi\psi}(\nabla_n\xi^r)(\nabla_m \xi_p)+\G_{tt}(\nabla_n\psi^r)(\nabla_m\psi_p)\\
&\hspace{3.5cm}-\G_{t\psi}\big(
	(\nabla_n\xi^r)(\nabla_m \psi_p)+(\nabla_n\psi^r)(\nabla_m \xi_p)
\big)\big]k_r\;,
 \end{aligned}
\end{equation}
and the fact that $k_r$ is arbitrary, we find that the Riemann tensor of $\CB$, denoted $\CR_{abcd}$ is given by
\begin{equation}
\begin{aligned}
 \mathcal R_{abcd}&=h_{[a}^{\phantom{[a} p}h_{b]}^{\phantom{b]} q}h_{[c}^{\phantom{[c} r}h_{d]}^{\phantom{d]} s}\Big\{
R_{pqrs}\\
&\hspace{3.3cm}-\frac{2}{\tau}\big[
	\G_{\psi\psi}(\nabla_p\xi_q)(\nabla_r \xi_s)+\G_{tt}(\nabla_p\psi_q)(\nabla_r \psi_s)\\
&\hspace{4.3cm}-\G_{t\psi}\big((\nabla_p\xi_q)(\nabla_r \psi_s)+(\nabla_p\psi_q)(\nabla_r \xi_s)\big)\big]\\
&\hspace{3.3cm}-\frac{2}{\tau}\big[
	\G_{\psi\psi}(\nabla_p\xi_r)(\nabla_q \xi_s)+\G_{tt}(\nabla_p\psi_r)(\nabla_q \psi_s)\\
&\hspace{4.3cm}-\G_{t\psi}\big((\nabla_p\xi_r)(\nabla_q \psi_s)+(\nabla_p\psi_r)(\nabla_q \xi_s)\big)\big]
\Big\}\;,
\end{aligned}
\end{equation}
where $R_{pqrs}$ is the Riemann tensor of the full spacetime. Contracting this expression with $h^{ac}$ and using \eqref{eqn:dxi} we finally obtain the desired equation%
\begin{equation}
 \mathcal R_{bd}=h_b^{\phantom b q}h_d^{\phantom d s}\,R_{qs}+\frac{1}{2}\,D_b\left(\frac{1}{\tau}\,D_d\tau\right)
	+\frac{1}{4}\,(\G^{-1})^{\alpha\mu}(D_b\G_{\alpha\beta})
		(\G^{-1})^{\beta\nu}(D_d\G_{\mu\nu})\,.
\label{eqn:ricci}
\end{equation}

Summarizing, the Einstein's equations for a spacetime with two commuting Killing vectors and invariant under $(t,\psi)\leftrightarrow (-t,-\psi)$ are given by equations \eqref{eqn:scalar} and \eqref{eqn:ricci}, with the Ricci tensor of the full spacetime $R_{mn}$ specified by the corresponding sources. In particular, for vacuum spacetimes we set $R_{mn} = 0$ in \req{eqn:scalar} and \req{eqn:ricci}, further simplifying the expressions.

Finally we note that  \reef{eqn:scalar} and \req{eqn:ricci} also hold for a spacetime with $N$ commuting Killing vectors when it admits a block diagonal metric with $\gamma$, now an $N \times N$ matrix (one block) and the other the base metric.

%__________________________________________________
\section{Construction of the double Myers-Perry solution}
\label{doubleMP}
%__________________________________________________

The double Myers-Perry solution discussed in \sec{ssdmp} is obtained by the Belinsky-Zakharov inverse scattering technique. We use the notation and nomenclature of \cite{Elvang:2007rd} (see also references therein).

There are two distinct Weyl solutions describing two static Myers-Perry black holes held apart by conical singularities. We consider only one of the configurations, but add angular momentum to it in the two different planes. The two different singly spinning cases are referred to as Configs.~A and B; see \fig{fig:ssdmp} for the rod diagrams. 

To obtain the solution of Config.~A, we start from the metric
\begin{equation}
 G_0=\diag\left\{-\frac{\mu_1\mu_3}{\mu_2\mu_4},\,\frac{\rho^2\mu_4}{\mu_1\mu_5},\,\frac{\mu_2\mu_5}{\mu_3}\right\}\,.
\end{equation}
We then remove an anti-soliton at $a_1$ and a soliton at $a_4$ from $(G_0)_{tt}$ and rescale the resulting metric by a factor of $\frac{\mu_1}{\mu_4}$. The solitons are then re-added using the BZ technique with vectors $(1,B_i,0)$, $i=1,4$ and finally we rescale back by a factor of $\frac{\mu_4}{\mu_1}$. The result is the desired metric. The BZ parameter $B_4$ must be fixed so that the two spacelike rods which were along the $\phi$-direction in the original static metric lie along the same direction after the BZ transformation. A coordinate change is required in order to bring the final metric into a manifestly asymptotically flat form. Note that in the final solution both black holes are spinning in the same plane and they each have intrinsic spin.
The solution cannot be balanced for any choice of parameters, so there are conical singularities in the metric. This, however, does not affect our results for the ergoregion mergers.

On the other hand, to generate the solution in Config.~B, we start from the metric
\begin{equation}
 G_0=\diag\left\{-\frac{\mu_0\mu_3}{\mu_2\mu_4},\,\frac{\rho^2\mu_4}{\mu_1\mu_5},\,\frac{\mu_1\mu_2\mu_5}{\mu_0\mu_3}\right\}\,.
\end{equation}
We remove an anti-soliton at $z=a_0$ from $(G_0)_{tt}$ and rescale the resulting metric by $\frac{\mu_0}{\rho^2}$. The anti-soliton is subsequently re-added with BZ vector $(1,0,C_0)$, and we rescale the resulting metric by a factor of $\frac{\rho^2}{\mu_0}$. Finally, we fix the BZ parameter $C_0$ to remove a naked singularity at $a_0$. The resulting solution has conical singularities, but is otherwise regular, and it is asymptotically flat. Again, the conical singularities do not influence our study of the ergoregion mergers. Note that in this solution both black holes are spinning along the $\psi$-direction, but only the left black hole carries a non-zero intrinsic (Komar) angular momentum. The other black hole rotates due to the frame dragging effect.

One can generalize the procedure above to obtain the solution where both Myers-Perry black holes in configuration B have independent angular momenta. 

%__________________________________________________
\section{Weyl coordinates for the doubly spinning ring}
\label{weyldspinring}
%__________________________________________________

We present here the coordinate transformation needed to go between ring coordinates $(x,y)$ and Weyl coordinates $(z,\rho)$ for the doubly spinning black ring.

Ref.~\cite{Pomeransky:2006bd} presented the expressions for $(z,\rho)$ in terms of $(x,y)$,
\begin{equation}
 \rho^2=-\frac{4\,G(x)G(y)}{(x-y)^4(1-\nu)^2}\,,\qquad 
z=\frac{(1-xy)\big[2+\lambda(x+y)+2\,\nu\,xy\big]}{(x-y)^2(1-\nu)}\,, 
\label{eqn:D2ringrhoz} 
\end{equation}
where $G(\xi)=(1-\xi^2)(1+\lambda\, \xi+\nu\,\xi^2)$, and the parameters $\lambda$ and $\nu$ satisfy the constraints \req{lamnu}. 
\fig{fig:D2ringrods} shows the rod structure of the doubly spinning black ring. The rod endpoints are at $z=\kappa_1$, $z=\kappa_2$ and $z=\kappa_3=1$, with $\kappa_{1,2}$  given in \req{D2kappa}.
For convenience, we work with dimensionless quantities, but the scale can be restored by taking $\rho \to k^2\,\rho$, $z \to k^2\, z$, $\kappa_i \to k^2 \kappa_i$. 

Using \eqref{eqn:D2ringrhoz} and \eqref{D2kappa}, we find that the functions $R_i=\sqrt{\rho^2+(z-z_i)^2}$, $i=1,2,3$, are simple expressions in terms of the $(x,y)$ coordinates:
\begin{subequations}
 \begin{align}
R_1&=\frac{1}{(x-y)(1-\nu )}\left[2+\sqrt{\lambda ^2-4 \nu }+\lambda (x+y)+\big(2\nu -\sqrt{\lambda ^2-4 \nu} \big)x\, y\right]\,,\\
R_2&=\frac{1}{(x-y)(1-\nu )}\left[2-\sqrt{\lambda ^2-4 \nu }+\lambda (x+y)+\big(2\nu +\sqrt{\lambda ^2-4 \nu} \big)x\, y\right]\,,\\
R_3&=\frac{1}{(x-y)(1-\nu )}\left[-\lambda -(1+\nu )(x+y)-\lambda \, x\, y\right]\;.
 \end{align}
\end{subequations}
These expressions can be inverted and we find
\begin{equation}
x=\frac{p(\rho,z)+c}{q(\rho,z)}\,,\qquad 
y=\frac{p(\rho,z)-c}{q(\rho,z)}\,,
\label{eqn:xyD2ring}
\end{equation}
where we have defined
\bea
\nonumber
 \hspace{-7mm}
 &&p(\rho,z)=\lambda (1-\nu )\sqrt{\lambda ^2-4 \nu }(R_1+R_2)-\lambda (1-\nu )^2(R_1-R_2)
	+2(1-\nu^2)\sqrt{\lambda^2-4 \nu }\,R_3\,,\\[1mm]
\nonumber
 \hspace{-7mm}
&&q(\rho,z)=-(1-\nu )\left[(1+\nu )\sqrt{\lambda ^2-4 \nu }(R_1+R_2)-\big(\lambda^2-2\nu (1+\nu )\big)(R_1-R_2)+2\lambda \sqrt{\lambda ^2-4 \nu }\, R_3\right]\,,\\[1mm]
 \hspace{-7mm}
&&c=2\,k^2\,\sqrt{\lambda ^2-4 \nu }\left(\lambda ^2-(1+\nu )^2\right)\,.
\label{eqn:pqrhoz}
\eea
The $R_i$'s should be regarded as functions of $\rho$ and $z$. Using \eqref{eqn:xyD2ring} with \eqref{eqn:pqrhoz} we can readily write the metric for the doubly spinning black ring in Weyl coordinates.

%___________________________________________________________
\section{Topology changes in ergosurface mergers}
\label{topology}
%___________________________________________________________

In this appendix we provide a simple and pictorial description of the topology changes that occur in the ergosurface mergers discussed in the main text. 

Let us first mention the general result of Hajicek \cite{Hajicek:1973fk}, who analyzed ergoregions in 3+1 dimensions, in particular for vacuum gravity. Hajicek showed that in 3+1 dimensions the ergosurface has to either touch the horizon at its poles or touch a singularity. The ergosurface will touch the horizon at so-called ``degenerate points", which are fixed points of the rotational isometry. It is easy to see how this result arises. For a stationary black hole, the horizon generators are of the form $\xi^a = (\partial_t)^a + \Omega_\phi\, (\partial_\phi)^a$. At the degenerate points, the Killing generator of the horizon $\xi^a$ reduces to just $(\partial_t)^a$. This suffices to ensure that the horizon and the ergosurface coincide locally. 

{}For singly spinning spherical black holes in higher dimensions, the ergosurface will touch the horizon at the fixed point of the rotational isometry. However, more generally the extension of the result of \cite{Hajicek:1973fk} to higher dimensions is less constraining. For multiply spinning objects the ergoregion would only touch the horizon where all the rotational isometries have fixed points simultaneously. Even for singly spinning solutions, the ergosurface need not touch the horizon: For instance, for a black ring with $\mathbf{S}^1$ angular momentum, the ergosurface does not touch the horizon anywhere because the rotational isometry (of the ring $\mathbf{S}^1$) does not have any fixed points on the horizon.

\begin{figure}[t!]
\begin{center}
 \includegraphics[scale=0.4]{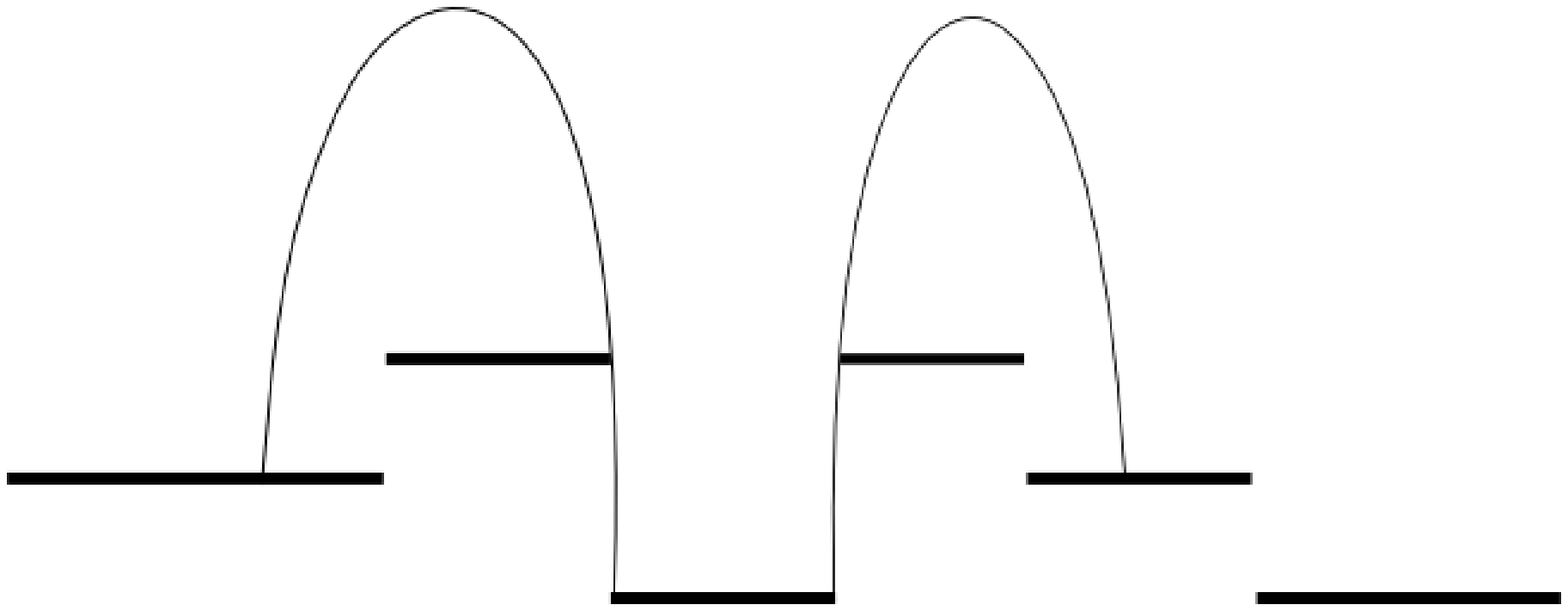}
\hspace{0.cm}
\raisebox{1.5cm}~~~~~~
\hspace{0.1cm}
 \includegraphics[scale=0.4]{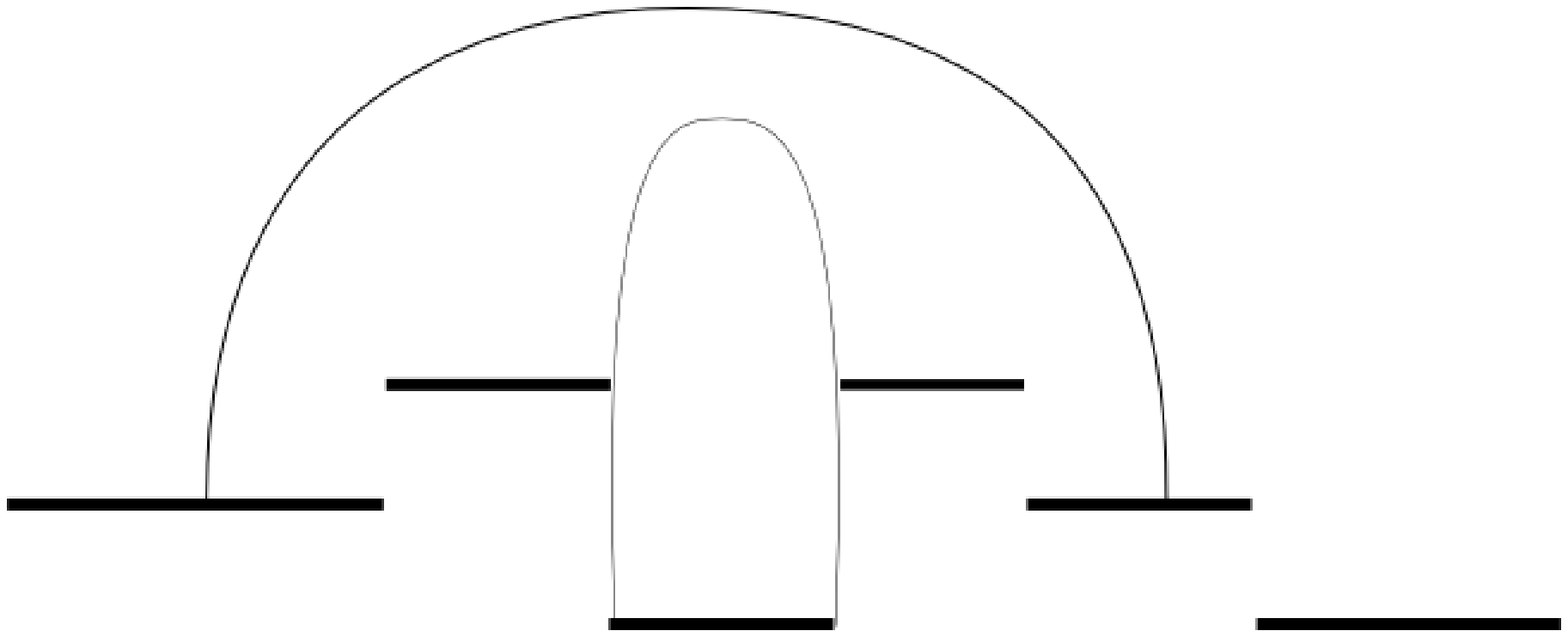}
\begin{picture}(0,0)
\setlength{\unitlength}{1cm}
%labels
\put(-16.8,1.0){\small{$t$}}
\put(-16.8,0.0){\small{$\psi$}}
\put(-16.8,0.5){\small{$\phi$}}
\put(-7.9,1.0){\small{$t$}}
\put(-7.9,0.0){\small{$\psi$}}
\put(-7.9,0.5){\small{$\phi$}}
%
%points
\put(-5.8,-0.5){\small{$a_1$}}
\put(-4.8,-0.5){\small{$a_2$}}
\put(-3.8,-0.5){\small{$a_3$}}
\put(-2.8,-0.5){\small{$a_4$}}
\put(-1.8,-0.5){\small{$a_5$}}
\put(-14.8,-0.5){\small{$a_1$}}
\put(-13.8,-0.5){\small{$a_2$}}
\put(-12.8,-0.5){\small{$a_3$}}
\put(-11.8,-0.5){\small{$a_4$}}
\put(-10.8,-0.5){\small{$a_5$}}
\put(-14.8,1.4){\tiny{$(1,0,\Omega_1)$}}
\put(-12.45,1.4){\tiny{$(1,0,\Omega_2)$}}
\put(-5.8,1.4){\tiny{$(1,0,\Omega_1)$}}
\put(-3.5,1.4){\tiny{$(1,0,\Omega_2)$}}
\end{picture}
\end{center}
%\vspace{1cm}
\caption{Topology change in ergosurfaces mergers for the singly spinning double Myers-Perry black hole configuration of type B.  Here we illustrate the topology change from two disjoint $\mathbf{S}^3$ ergosurfaces (left) to an outer and an inner ergosurface, each of topology $\mathbf{S}^2\times\mathbf{S}^1$ (right).}
\label{fig:topology}
\end{figure}

Ergosurfaces are conveniently represented in the $(z,\rho)$ plane for generalized Weyl solutions, but to visualize their topology, it is useful to represent them schematically in the rod diagram. Recall that the rod diagram represents the various segments (rods) of the $z$-axis where a particular Killing field vanishes. Different rods are drawn at different heights to indicate where each Killing vector vanishes.\footnote{
One has to keep in mind though that for rotating black holes, the horizon rod does not lie along asymptotic timelike Killing field, $(\partial_t)^a$.} If we sketch the ergosurfaces as curves on the rod diagram, they will end on different rods according to their topology.  

An example is given in \fig{fig:topology} which sketches the ergosurfaces of the double Myers-Perry solution of type B (see section \sec{ssdmp}). On the left, the two black holes are far apart and each of them has an ergosurface with $\mathbf{S}^3$ topology: in the rod diagram endpoints of each ergosurface curve lie on rods for different rotational directions. At such a point, the corresponding $\mathbf{S}^1$ closes off, but it has finite radius anywhere else on the ergosurface. The two rotational Killing vectors vanish at different points, so the topology is $\mathbf{S}^3$. 
Since this configuration is spinning in a single plane, the ergosurfaces are pinned at the horizons only at poles where the corresponding Killing vector vanishes. 
The configuration on the right of \fig{fig:topology} shows the ergosurfaces after the merger. In this case, there is an outer and an inner ergosurface, each of topology $\mathbf{S}^2\times\mathbf{S}^1$. Now the inner ergosurface is pinned at the poles of the horizons.

%%%%%%%%%%%%%%%%%%%%%%%%%%%%%%%%%%%%%%%%%%%%
%\bibliographystyle{utphys}
%\bibliography{ergo}

\providecommand{\href}[2]{#2}\begingroup\raggedright\endgroup

\end{document}